\documentclass[12pt,reqno]{amsart}
\usepackage{amsmath,amssymb,amsfonts,amsthm}
\usepackage[mathscr]{eucal}
\usepackage[all]{xy}
\usepackage{hyperref}
\usepackage{setspace}
\usepackage{bm}
\usepackage{xcolor}
\allowdisplaybreaks

\author{V.~A.~Abakumova, I.~Yu.~Karataeva, S.~L.~Lyakhovich}

\address{Physics Faculty, Tomsk State University, Lenin ave. 36, Tomsk 634050, Russia.}

\email{abakumova@phys.tsu.ru, \, karin@phys.tsu.ru,\, sll@phys.tsu.ru}

\title[Reducible gauge symmetry versus unfree gauge symmetry ...]{Reducible gauge symmetry versus unfree gauge symmetry \\ in Hamiltonian formalism}

\begin{document}

\begin{abstract}
The unfree gauge symmetry implies that gauge variation of the action functional vanishes provided for the gauge parameters are restricted by the differential equations. The unfree gauge symmetry is shown to lead to the global conserved quantities whose on shell values are defined by the asymptotics of the fields or data on the lower dimension surface, or even at the point of the space-time, rather than Cauchy hyper-surface. The most known example of such quantity is the cosmological constant of unimodular gravity. More examples are provided in the article for the higher spin gravity analogues of the cosmological constant. Any action enjoying the unfree gauge symmetry is demonstrated to admit the alternative form of gauge symmetry with the higher order derivatives of unrestricted gauge parameters. The higher order gauge symmetry is reducible in general, even if the unfree symmetry is not. The relationship is detailed between these two forms of gauge symmetry in the constrained Hamiltonian formalism. The local map is shown to exist from the unfree gauge algebra to the reducible higher order one, while the inverse map is non-local, in general. The Hamiltonian BFV-BRST formalism is studied for both forms of the gauge symmetry. These two Hamiltonian formalisms are shown connected by canonical transformation involving the ghosts. The generating function is local for the transformation, though the transformation as such is not local, in general. Hence, these two local BRST complexes are not quasi-isomorphic in the sense that their local BRST-cohomology groups can be different. This difference in particular concerns the global conserved quantities. From the standpoint of the BRST complex for unfree gauge symmetry, these quantities are BRST-exact, while for the alternative complex, these quantities are the non-trivial co-cycles.
\end{abstract}

\maketitle
%\vspace{-5mm}
%\begin{center}
%\emph{E-mail adresses:}\,\,{abakumova@phys.tsu.ru}, {sll@phys.tsu.ru}\\
%\emph{Physics Faculty, Tomsk State University, Lenin ave. 36, Tomsk 634050, Russia}
%\end{center}

\section{Introduction}
Gauge symmetry transformations are defined by the gauge
parameters being functions of the space-time point $x$. It is
assumed that the gauge variation of the fields can change the action
functional at most by the surface terms. Unfree gauge variation
implies that the gauge parameters are not arbitrary functions of
$x$, but are solutions to a system of partial differential equations
(PDE). The most known example of the unfree gauge symmetry is the
volume-preserving diffeomorphism of the unimodular gravity (for
review and further references, see \cite{Percacci2018},
\cite{Gielen2018AP}). This gauge transformation is parameterized by
the vector field $\epsilon^\mu(x)$ which is constrained by the
transversality equation $\nabla_\mu\epsilon^\mu=0$. The
transversality  condition is also imposed on the gauge parameter of
the Fierz-Pauli spin two field theory \cite{Alvarez:2006uu},
\cite{Blas:2007pp}. There are several analogues among the higher
spin field theories \cite{SKVORTSOV2008301}, \cite{Campoleoni2013}
where the tensorial gauge parameters are constrained by PDE's. It is
noticed in the article \cite{FRANCIA2014248}, that the linear field
theories with unfree gauge symmetry proposed in the articles
\cite{Alvarez:2006uu}, \cite{Blas:2007pp}, \cite{SKVORTSOV2008301},
\cite{Campoleoni2013} admit an alternative parametrization of gauge
transformations.  The alternative transformations are reducible
(i.e. there is symmetry of symmetry) and they involve the higher
order derivatives of gauge parameters. Furthermore, it is proven
that at free level any local field theory always admits
parametrization of gauge transformations by unconstrained gauge
parameters with finite reducibility order \cite{FRANCIA2014248}.
Thus, at least at the level of linear(ized) field equations, any
model with unfree gauge symmetry can be alternatively treated as the
system with the reducible gauge symmetry.

In the article \cite{KAPARULIN2019114735}, the general structure of
unfree gauge algebra is described. The key source of distinctions
from the case with unconstrained gauge parameters is that the second
Noether theorem does not apply once the gauge parameters are
restricted by PDE's. This distinction results in modification of the
Noether identities which involve, besides the gauge generators, the
operators of the gauge parameter constraints. Proceeding from the
modified gauge identities, the Faddeev-Popov quantization rules are
adjusted to properly account for the equations imposed on the gauge
parameters \cite{KAPARULIN2019114735}. In the article
\cite{Kaparulin2019}, the Batalin-Vilkovisky (BV) field-anti-field
formalism is  worked out for the systems with unfree gauge symmetry.
The articles \cite{Abakumova:2019uoo}, \cite{Abakumova:2020ajc},
\cite{Abakumova:2021rlq} describe the phenomenon of unfree gauge
symmetry in terms of constrained Hamiltonian formalism. In
particular, even in the well-known Hamiltonian formalism of
unimodular gravity \cite{Unruh1989}, \cite{Henneaux1989T}, it
remained not evident how the transversality condition for gauge
parameters reveals itself at the level of constraint algebra, as
noted in the review \cite{Gielen2018AP}. The articles
\cite{Abakumova:2019uoo}, \cite{Abakumova:2020ajc} identify the
structures  of the Hamiltonian involution relations which encode the
differential equations imposed on the gauge parameters. The works
\cite{Abakumova:2019uoo}, \cite{Abakumova:2020ajc},
\cite{Abakumova:2021rlq} also provide the
BFV-BRST\footnote{Batalin-Fradkin-Vilkovisky---Becchi-Rouet-Stora-Tyutin}
Hamiltonian formalism for the systems with unfree gauge symmetry.

One of the most attractive features of the unimodular gravity is that
cosmological constant $\Lambda$ is  the ``global conserved
quantity'', see \cite{Percacci2018} for discussion and further
references. In the unimodular gravity, and in the recent
modifications \cite{BARVINSKY201759}, \cite{Barvinsky-Kolganov},
$\Lambda$ appears as the integration constant whose specific value
is defined by the asymptotics of the fields, unlike Einstein's
gravity where $\Lambda$ is the predefined parameter included into
the Lagrangian. It is noted in the articles
\cite{KAPARULIN2019114735}, \cite{Kaparulin2019}, that existence of
such integration constants is a common feature for all the systems
with unfree gauge symmetry. These ``global conserved quantities''
are found to be a direct consequence of the differential equations
restricting gauge parameters. Once the specific values of these
integration constants are defined by field asymptotics, and not by
initial data, they are considered as the modular parameters of the
theory, rather than the local conserved quantities like the currents. The examples of the
higher spin analogues for cosmological constant are noted in the
articles \cite{Abakumova:2020ajc}, \cite{Abakumova:2021rlq}. For
instance, for the spin three model with the transverse gauge
symmetry proposed in \cite{SKVORTSOV2008301}, there are fifteen
``global conserved quantities" in $d=4$ \cite{Abakumova:2020ajc}.
Another specific example of the higher spin analogues of the
unimodular gravity cosmological constant is provided in the next
section.

In this article, we demonstrate at the level of constrained
Hamiltonian formalism that any system with unfree gauge symmetry
always admits an alternative description in terms of reducible gauge
symmetry with the higher order derivatives of the gauge parameters.
The higher derivative gauge transformations are known for a long
time, since the very early years of the constrained Hamiltonian
formalism, \cite{Anderson1951B}, \cite{Utiyama1959},
\cite{Mukunda1980}, \cite{Castellani1982}. The basic origin of these
 symmetries is  connected with the secondary constraints: if the
Dirac-Bergmann algorithm terminates without defining the Lagrange
multipliers, this means that the time derivatives of the final stage
constraints reduce to the combinations of previous ones. In its own
turn, this means the Noether identity between the equations. Once
the secondary constraints arise as the differential consequences of the
primary ones and Hamiltonian equations, the Noether identity
generator involves the second time derivative. Tertiary constraints
mean the third derivatives of the equations involved in the Noether
identities, etc. By the second Noether theorem, the gauge symmetry
transformations should involve the same order time derivatives of
the gauge parameters as the order of the Noether identity. As first
noted in the article \cite{KAPARULIN2019114735}, the unfree gauge
symmetry implies certain modification of the second Noether theorem,
to account for the equations constraining gauge parameters. This
modification is briefly presented in the next section. As further
explained in this section, the modified gauge identities can be
converted into the usual ones. In general, the latter Noether
identities are of a higher order, and they are reducible. Hence, the action
functional enjoying the unfree gauge symmetry always possesses the
usual reducible gauge symmetry of a higher order. For linear field
theories, this fact has been previously noted in the article
\cite{FRANCIA2014248}. In this article, we extend it to the general
constrained Hamiltonian systems. As the same constrained Hamiltonian
system admits two alternative forms of gauge symmetry, the question
arises about their interconnection. This issue is clarified in the
end of Section \ref{Section3} of the article. Both reducible and  unfree form of
the gauge symmetry can be connected with corresponding BRST complex.
Therefore, the two BRST complexes turn out connected with the same
original system. The relation between these complexes is clarified
in Section \ref{Section4}: they are connected by the canonical transformation
with the local generating function, though the transformation, as such, is
non-local. In general, these two local BRST complexes are not
quasi-isomorphic in the sense that their local BRST cohomology groups can be
different. The subtle difference is due to the way of describing the
modular parameters in terms of these complexes. The BRST complex
which corresponds to the reducible form of the gauge symmetry has a
cohomology group related to the modular parameters. The complex
based on the unfree gauge symmetry does not have similar group,
while the modular parameters are explicitly involved in the BRST
operator as the fixed constants. This subtlety will be further studied
elsewhere. In Section \ref{Section5}, the general formalism is exemplified by the model of
linearized unimodular gravity.

\section{Unfree and reducible gauge symmetry in Lagrangian formalism}\label{Section2}

In this section, we briefly outline
 the basic generalities of unfree gauge symmetry in Lagrangian setup, for rigorous exposition see \cite{KAPARULIN2019114735}, \cite{Kaparulin2019}. The issue of the ``global conserved quantities'' is addressed in more details than in earlier works \cite{KAPARULIN2019114735}, \cite{Kaparulin2019}. In particular, the specific procedure is proposed to compute these conserved quantities, given the unfree gauge symmetry.
 Then, we explain the general connection between the unfree gauge symmetry and the reducible one, and exemplify the general formalism by specific models.
 In the next section, this connection between two forms of the gauge symmetry is explicitly realized for general  constrained Hamiltonian systems.

Consider the set of fields\footnote{In this section, we use the condensed notation, besides discussion of the examples.
The condensed indices include the integer labels and the space-time point $x$. Summation over the condensed indices includes the integral by $x$.
The derivatives $\partial_i=\frac{\partial}{\partial\phi^i}$ are understood as variational. For details of applying the condensed notation, see \cite{DeWitt:1965jb}, or \cite{teitelboim1992quantization}.} $\phi^i$ with the action functional $S(\phi)$. Let the Lagrangian equations
\begin{equation}\label{LE}
  \partial_i S(\phi)=0\,
\end{equation}
obey the identities
\begin{equation}\label{GI}
  \partial_i S(\phi)\Gamma^i{}_\alpha (\phi)+\tau_a(\phi)\Gamma^a{}_\alpha(\phi)\equiv 0\,.
\end{equation}
Here $\Gamma^i{}_\alpha (\phi)$ and $\Gamma^a{}_\alpha (\phi)$ are the matrices of the local differential operators,
and $\tau_a(\phi)$ are the functions of the fields and their space-time derivatives.
If $\tau_a(\phi )$ and/or $\Gamma^a{}_\alpha (\phi)$ reduced to the linear combination of $\partial_i S(\phi)$,
relations (\ref{GI}) would present the Noether identities between Lagrangian equations (\ref{LE}).
This would lead to the gauge symmetry of the action with unconstrained gauge parameters.
The case of unfree gauge symmetry implies non-trivial $\Gamma^a{}_\alpha (\phi)$ and $\tau_a(\phi)$.
In particular, $\Gamma^a{}_\alpha (\phi)$ is supposed to be a differential operator with a  finite kernel,
$\dim\ker\Gamma^a{}_\alpha (\phi)=k\in\mathbb{N}$.
This means, that the space of solutions is finite-dimensional for the equations
\begin{equation}\label{KerG}
  u_a\Gamma^a{}_\alpha(\phi)=0 \quad \Rightarrow \quad
  u_a\in K\,, \quad \dim K= k\in\mathbb{N}\,.
\end{equation}
As the equations above are linear, the space of solutions $K$ is a linear space.
Choosing the basis $\{u^\mathcal{I}{}_a,\, \mathcal{I}=1,\dots , k\}$, we can span any solution
 $u_a\in K \,\Rightarrow\, u_a=\Lambda_\mathcal{I}u^\mathcal{I}{}_a$, where $\Lambda_\mathcal{I}$ are the constants.
 These constants are considered as the modular parameters which allow one to distinguish different classes of solutions
 of Lagrangian equations (\ref{LE}). This issue will be commented later in this section.
 Given the property (\ref{KerG}), one can ``integrate" equations (\ref{GI}) with respect to
  the quantities $\tau_a(\phi)$:
\begin{equation}\label{tau}
  \mathcal{T}_a(\phi,\Lambda)\equiv\tau_a(\phi)-\Lambda_\mathcal{I}u^\mathcal{I}{}_a\approx 0\,,
\end{equation}
 where $\approx$ stands for the on-shell equality. Once the basis of solutions $u^\mathcal{I}$ is supposed independent, all the constants
 $\Lambda_\mathcal{I}$ are supposed to be essentially involved in the
 equations above.
 This means, the relations (\ref{tau}), and their
 differential consequences can be resolved with respect to the
 modular parameters $\Lambda_\mathcal{I}$:
\begin{equation}\label{J}
  \Lambda_\mathcal{I}\approx J_\mathcal{I}(\phi)\,.
\end{equation}
In its own turn, this means that $k=\dim \text{Ker}\,\Gamma^a{}_\alpha$
independent local quantities exist that remain constant on shell.
These ``conserved quantities'' $J_\mathcal{I}(\phi)$ are distinct from the
conserved charges, because no conserved currents are connected to
them. Furthermore, unlike the conserved charge, the quantity
$J_\mathcal{I}(\phi)$ is not an integral over any $(d-1)$-dimensional
hyper-surface in the space-time  --- it is just a function of the
fields and their derivatives. This function remains a constant on
shell, i.e. it does not change under variation of the space-time
arguments of the fields. This means, the specific numerical value
$\Lambda_\mathcal{I}$ of $J_\mathcal{I}(\phi)$ is defined by the fields and their
derivatives at asymptotics, or at a single space-time point.  No
transition is admissible at classical level between the states with
different values of $J_\mathcal{I}(\phi)$.  So,  the variety of solutions of
Lagrangian equations is decomposed into the classes with different
modular parameters, corresponding to different asymptotics.
 In the simplest case of the fields vanishing at infinity, all the modular parameters $\Lambda$ obviously
 vanish.
 So, if the consideration is restricted by the fields with vanishing asymptotics, the local quantity $\tau_a(\phi)$ will vanish on shell, while it does not reduce to a linear combination of the l.h.s. of Lagrangian equations. In this sense, the mass shell, being a zero locus of $\partial_i S(\phi)$ is incomplete, as the local quantities $  \mathcal{T}_a(\phi,\Lambda)$ are on-shell trivial, while $\mathcal{T}_a(\phi,\Lambda)\neq\partial_i S\,\Theta^i{}_a$. With this regard, the local quantities $\tau_a$ are termed as the mass shell completion functions. Any on-shell trivial local quantity is supposed to be the linear combination of $\partial_i S$ and $\tau_a(\phi)$.
 The local quantities  are considered trivial, if they reduce on shell to the field independent expressions.
 The l.h.s. of the Lagrangian equations (\ref{LE}) and the completion functions (\ref{tau}) are supposed to constitute the generating set for the variety of trivial quantities. Relations (\ref{GI}) between the elements of the generating set for on shell vanishing variety of local quantities are considered as modified Noether identities.

Once the  Noether identities are  modified (\ref{GI}), this results
in the unfree gauge symmetry. The gauge transformations read
\begin{equation}\label{Unfree}
  \delta_\epsilon\phi^i=\Gamma^i{}_\alpha(\phi)\epsilon^\alpha\,.
\end{equation}
These gauge transformations leave the action functional off-shell,
invariant
\begin{equation}\label{GSS}
  \delta_\epsilon S(\phi)\equiv 0\,,
\end{equation}
 for any gauge parameters $\epsilon^\alpha$ that obey
equations
\begin{equation}\label{GPE}
  \Gamma^a{}_\alpha(\phi)\epsilon^\alpha=0\,.
\end{equation}
The operators $\Gamma^a{}_\alpha (\phi)$, being involved in the
modified Noether identities (\ref{GI}), and having the
finite-dimensional kernel (\ref{KerG}), turn out defining the
equations (\ref{GPE}) imposed on the gauge parameters. These
equations make the gauge variation (\ref{Unfree}) unfree. Given the
equations (\ref{GPE}), we term $\Gamma^a{}_\alpha$ as the operators of
gauge parameter constraints. The operators $\Gamma^i{}_\alpha$, being
involved in the modified Noether identities (\ref{GI}) as the
coefficients at Lagrangian equations, obviously play the role of the
generators of unfree gauge symmetry.

If the fields are defined in $d$-dimensional space-time, then the
general solution of equations (\ref{GPE}) must include arbitrary
functions of $d$ space-time coordinates, otherwise it is not gauge
symmetry. This means, the linear equations (\ref{GPE}) for
$\epsilon^\alpha$ admit a basis of solutions parameterized by
arbitrary functions $\mathcal{E}^A$, where $A$ is a condensed index
including the space-time coordinates,
\begin{equation}\label{rho}
  \Gamma^a{}_\alpha(\phi)\epsilon^\alpha=0 \quad \Leftrightarrow \quad
  \exists\,\rho^\alpha{}_A(\phi): \quad \epsilon^\alpha=\rho^\alpha{}_A(\phi)\mathcal{E}^A\,,
  \quad \forall\,\mathcal{E}^A\,.
\end{equation}
Contracting the modified identities (\ref{GI}) with $\rho^\alpha{}_A$,
 we arrive at the usual Noether identities
\begin{equation}\label{NI}
  \partial_i S(\phi)R^i{}_A(\phi)\equiv 0\,, \quad
  R^i{}_A(\phi)=\Gamma^i{}_\alpha(\phi)\rho^\alpha{}_A(\phi)\,.
\end{equation}
By virtue of these identities, the gauge transformations generated
by $ R^i{}_A(\phi)$ with arbitrary gauge parameters $\mathcal{E}^A$
leave the action invariant,
\begin{equation}\label{GSR}
  \delta_{\mathcal{E}}\phi^i=R^i{}_A(\phi)\mathcal{E}^A,\, \quad
  \delta_{\mathcal{E}} S(\phi)\equiv 0\,, \quad \forall\,\mathcal{E}^A\,.
\end{equation}
So we see that, given the unfree gauge symmetry (\ref{Unfree}),
(\ref{GSS}), (\ref{GPE}), one always arrives at the gauge symmetry
(\ref{rho}), (\ref{NI}), (\ref{GSR}) with unconstrained parameters,
while the action is the same.

Once  $\rho^\alpha{}_A$ (\ref{rho}) are the matrices of differential
operators, in general, the transformations with unconstrained
parameters (\ref{NI}), (\ref{GSR}) can involve the higher order
derivatives than the unfree gauge transformations (\ref{Unfree}),
(\ref{GPE}). One more remark is that the gauge generators
$R^i{}_A$ (\ref{NI}) can be reducible even if the original
generators $\Gamma^i{}_\alpha$ of the unfree gauge symmetry are
independent.

An important notice is that any gauge invariant of the unfree gauge
symmetry is also invariant of the reducible gauge symmetry, because
of relations (\ref{rho}), (\ref{NI}) connecting them. There is a
subtlety, however, related to the completion functions (\ref{tau}).
These quantities are on shell invariant under the unfree gauge
transformations \cite{KAPARULIN2019114735}, and hence, they are
invariant w.r.t. the transformations with the unconstrained
parameters (\ref{GSR}). As the completion functions do not reduce to
the linear combinations of Lagrangian equations, they are considered
to be non-trivial gauge invariants w.r.t the transformation
(\ref{GSR}). Hence, they would give rise to the non-trivial BRST
co-cycles if the gauge symmetry is described by corresponding
complex. From the standpoint of unfree gauge symmetry, the
completion functions (\ref{tau}) are treated as trivial quantities
as they are involved into the Noether identities (\ref{GI}) corresponding to
this symmetry. Corresponding BRST complex treats these quantities as
BRST co-boundary both in Lagrangian \cite{Kaparulin2019} and
Hamiltonian \cite{Abakumova:2019uoo}, \cite{Abakumova:2020ajc}
framework. This discrepancy is due to the fact that operator
$\rho^\alpha{}_A$, which maps unfree gauge transformations to the
unconstrained higher order ones, is not invertible, in general. This
distinction will be detailed in the Hamiltonian formalism in the
next section.

Below, we exemplify the general notions introduced in this section
by two field theory models.

\subsection*{Example 1: Unimodular gravity.}
Consider the unimodular gravity action\footnote{We use the following definitions for the Riemann tensor, Ricci tensor, and scalar curvature:
$ R^\alpha{}_{\beta\mu\nu}=\partial_\mu\Gamma^\alpha_{\nu\beta}-\partial_\nu\Gamma^\alpha_{\mu\beta}+\Gamma^\alpha_{\mu\gamma}\Gamma^{\gamma}_{\nu\beta}-\Gamma^\alpha_{\nu\gamma}\Gamma^{\gamma}_{\mu\beta}\,, \,
R_{\alpha\beta}=R^\gamma{}_{\alpha\gamma\beta}, \,R=g^{\alpha\beta}R_{\alpha\beta}\,.
$ In this subsection, Greek indices take the values $0,1,2,3$.} in $d=4$,
\begin{equation}
\text{det}\,{g_{\alpha\beta}}=-1\,, \quad S[g(x)]=-\int d^4x\,R\, .
\end{equation}
The Lagrangian equations read
\begin{equation}\label{UGLE}
     -\frac{\delta S}{\delta g^{\alpha\beta}}\equiv
R_{\alpha\beta}-\frac{1}{4}g_{\alpha\beta}R=0\, .
\end{equation}
Unlike the case of Einstein's equations, the divergence of equations
(\ref{UGLE}) does not identically vanish, but reduces to the partial
 derivative of scalar curvature. This is the case of the modified
 Noether identity (\ref{GI}):
\begin{equation}\label{UGGI}
\displaystyle -2\,\nabla^\beta\frac{\delta S}{\delta
g^{\alpha\beta}}-\partial_{\alpha}\tau=0\,, \quad \tau\equiv \frac{1}{2}R \, .
\end{equation}
The role of the operator $\Gamma^a{}_\alpha (\phi)$ of (\ref{GI}) is
played by $\partial_\alpha$, and $\frac{1}{2}R$ is recognized
as the completion function $\tau_a(\phi)$. Obviously, the kernel of
$\partial_\alpha$ is one-dimensional, being constituted by arbitrary
constants,
\begin{equation}
\displaystyle \partial_\alpha u=0 \quad \Rightarrow \quad u=\Lambda\,,
\quad \forall\,\Lambda\in \mathbb{R} \, ,
\end{equation}
cf. (\ref{KerG}).

Relation (\ref{tau}) for the completion function in the case of
unimodular gravity reads
\begin{equation}\label{TgLambda}
\displaystyle \mathcal{T}(g,\Lambda)\equiv \frac{1}{2}R-\Lambda\approx 0\,, \quad
 \forall\, \Lambda\in\mathbb{R} \, .
\end{equation}
In this case, we have single completion function $\tau$
(\ref{tau}) proportional to scalar curvature $R$. The  conserved
quantity $J$ (\ref{J}) in this case coincides with $\tau$. This coincidence
is due to the fact that the kernel of the operator $\partial_\alpha$ is
one-dimensional. In the next example, the completion functions
differ from conserved quantities.

Given the modified Noether identities (\ref{GI}) for any specific
model, one can  recognize  both the generator of unfree gauge
symmetry transformations (\ref{Unfree}) and the operator of the
equations imposed on the gauge parameters (\ref{GPE}). As the
modified Noether identities are found for the unimodular gravity
(\ref{UGGI}), we apply the general recipe (\ref{Unfree}),
(\ref{GPE}). The coefficient at Lagrangian equations in the identity
(\ref{UGGI}) should define the gauge transformations. Once it is the
same operator as in the gauge identity of the General Relativity,
the gauge transformation is the same:
\begin{equation}\label{diff}
    \delta_\epsilon g_{\alpha\beta}=-\,\nabla_\alpha\epsilon_\beta-
    \nabla_\beta\epsilon_\alpha \, .
\end{equation}
The coefficient at the completion  function $\tau$ in the identity
(\ref{UGGI}) defines the equation imposed on the gauge parameter
according to the recipe (\ref{GPE}). Hence, the equation (\ref{GPE})
imposed onto the gauge parameters, for the unimodular gravity reads:
\begin{equation}\label{UGTdiff}
\partial_\alpha\epsilon^\alpha=0 \, .
\end{equation}
Now, let us turn to identifying the higher order gauge symmetry with unconstrained parameters for
the unimodular gravity. Following the
general recipe, we have to begin with identifying the null-vectors
$\rho^\alpha{}_A$ (see relation (\ref{rho})) for the operators
$\Gamma^a{}_\alpha$ being the coefficients at completion functions in
the modified Noether identities (\ref{GI}). In the case of
unimodular gravity, the term with completion function $\tau=\frac{1}{2}R$ in
the modified Noether identity  (\ref{UGGI}) reads as
$-\,\partial_\alpha\tau$. Once it is the exact one-form, the null-vector is
the de Rham differential. Taking the differential of the l.h.s. of
the modified gauge identity (\ref{UGGI}), we arrive at the second
order Noether identity (cf. (\ref{NI})) for unimodular gravity
\begin{equation}\label{UGNI}
\displaystyle \big(\delta^\alpha_\nu\partial_\mu-\delta_\mu^\alpha\partial_\nu\big)\nabla^\beta\frac{\delta S}{\delta g^{\alpha\beta}}\equiv 0 \, .
\end{equation}
By the second Noether theorem, this identity means the
infinitesimal gauge symmetry of the action (\ref{UGLE}) with the gauge parameter, being antisymmetric tensor
\begin{equation}\label{UGRS}
\displaystyle
\delta_{\mathcal{E}}g_{\alpha\beta}=-\,g_{\beta\mu}\nabla_\alpha\partial_\nu\mathcal{E}^{\nu\mu}-
g_{\alpha\mu}\nabla_\beta\partial_\nu\mathcal{E}^{\nu\mu} \, , \qquad \mathcal{E}^{\alpha\beta}= -\, \mathcal{E}^{\alpha\beta}\, .
\end{equation}
This transformation is obviously reducible, as the gauge parameter $\mathcal{E}^{\alpha\beta}$ admits the gauge symmetry of its own,
\begin{equation}\label{UGGS2}
  \delta_\omega\mathcal{E}^{\alpha\beta}=\partial_\gamma\omega^{\alpha\beta\gamma}\, ,\qquad \omega^{\alpha\beta\gamma} = -\,\omega^{\beta\alpha\gamma} =
  -\,\omega^{\alpha\gamma\beta} \, .
\end{equation}
The parameter of this gauge symmetry is gauge invariant again,
\begin{equation}\label{UGGS3}
\delta_\eta\omega^{\alpha\beta\gamma}=\varepsilon^{\alpha\beta\gamma\lambda}\partial_\lambda \eta \, ,
\end{equation}
where $\varepsilon^{\alpha\beta\gamma\lambda}$ is the Levi-Civita symbol. The scalar gauge parameter $\eta$ is irreducible, so the sequence of the symmetry for symmetry transformations terminates at the second step.

The transformation (\ref{UGRS}) is the minus Lie derivative of the metrics with respect to the vector field $\epsilon^\alpha$ being the divergence of the antisymmetric tensor $\mathcal{E}^{\alpha\beta}$,
\begin{equation}\label{UGTdiffE}
\delta_{\mathcal{E}}g_{\alpha\beta}=-\,\nabla_\alpha\epsilon_\beta-\nabla_\beta\epsilon_\alpha\, , \qquad
\epsilon^\alpha=\partial_\nu\mathcal{E}^{\nu\alpha}\, .
\end{equation}
Obviously, the relation
$\epsilon^\alpha=\partial_\nu\mathcal{E}^{\nu\alpha}$ gives the general
local solution to the transversality condition (\ref{UGTdiff}).
Globally, the transverse vectors can exist such that do not reduce
to the divergence of any antisymmetric tensor. This depends on the
de Rham cohomology group of the space-time.

The parameterization of the volume preserving diffeomorphisms by the
antisymmetric tensors (\ref{UGTdiffE}) is long known
\cite{Dragon1988K}, \cite{Alvarez2013H}. In particular, in the
article \cite{Alvarez2013H}, it has been noted  that non-trivial
Betti numbers can matter if one opts to this parameterization. Let
us elaborate on this issue and its consequences for the
(in)equivalence between the unfree gauge transformations
(\ref{diff}), (\ref{UGTdiff}) and their reducible counterparts
(\ref{UGRS}). Consider the three-form $\tilde\epsilon=*\epsilon$
being the Hodge dual of vector $\epsilon$. The vector is transverse,
iff the form is closed $d\tilde\epsilon=0$. If $\tilde\epsilon$  is
exact, i.e.
$\exists\,\tilde{\mathcal{E}}:\,\tilde{\epsilon}=d\tilde{\mathcal{E}}
$, then the divergence of the bi-vector
$\mathcal{E}=*\tilde{\mathcal{E}}$ will give the original vector
$\epsilon$. If the closed non-exact three-forms exist, they give
rise to the transverse vectors that do not reduce to the divergence
of any bi-vector. So, the unfree gauge symmetry (\ref{diff}),
(\ref{UGTdiff}) is ``stronger'', in a sense, than the reducible
gauge symmetry (\ref{UGRS}). The unfree gauge symmetry gauges out
degrees of freedom connected with the de Rham cohomology group,
while the reducible gauge symmetry treats these global modes as
gauge invariants. This distinction between these two symmetries of
the same action can manifest itself in the subtleties related to the
topology of the space-time. Also notice that the symmetry for
symmetry transformations (\ref{UGGS2}) mean that the antisymmetric
tensor $\mathcal{E}^{\alpha\beta}$ is defined modulo transverse
second-rank tensor. The latter is not necessarily of the form
(\ref{UGGS2}) if the second de Rham cohomology group is non-trivial.
So,  one can opt for treating the gauge as unfree, with the higher
symmetry parameters restricted by transversality conditions.
Corresponding BV-BRST complex can be constructed along the lines of
the work \cite{Kaparulin2019}, with the constraints imposed on the
ghosts. The complex related to the reducible gauge symmetry will
differ from the one associated to the unfree gauge symmetry by the
cohomology groups in the positive ghost numbers. These distinctions,
originating from de Rham cohomology, are connected with topology of
the space-time. For the local considerations, the
only distinction is the interpretation of the completion function
$\tau$ (\ref{UGGI}),  and corresponding modular parameter $\Lambda$,
being the cosmological constant. For reducible gauge symmetry
(\ref{UGRS}), $\tau$ in (\ref{UGGI}), (\ref{TgLambda}) is a non-trivial observable.
 At the level of BV formalism, it would correspond to BRST co-cycle. For the unfree
gauge symmetry, $\tau$ in (\ref{UGGI}), (\ref{TgLambda})
would be BRST exact, while $\Lambda$ is a constant
parameter included into the BRST operator.

\subsection*{Example 2: Maxwell-like theory of the third rank tensor field}
This example mostly follows the pattern of the previous one, while
the main distinctions concern the conserved quantities (\ref{J})
related to the kernel of the operator $\Gamma^a{}_\alpha$ of the
equation (\ref{GPE}) constraining gauge parameters. In the case of
unimodular gravity, single equation (\ref{UGTdiff}) leads to one
conserved quantity -- cosmological constant. The model of this
section has four equations constraining gauge parameters. These
four equations lead to ten conserved quantities.

Consider Lagrangian for the tracefull third-rank symmetric tensor
field proposed in Ref. \cite{Campoleoni2013}
\begin{equation}\label{M-LS}
\displaystyle
\mathcal{L}=-\,\frac{1}{2}\big(\partial_\lambda\varphi_{\alpha\beta\gamma}\partial^\lambda\varphi^{\alpha\beta\gamma}-3\partial^\alpha\varphi_{\alpha\beta\gamma}\partial_\lambda\varphi^{\lambda\beta\gamma}\big)\,,
\end{equation}
which describes  massless spin one and three. The field equations
read
\begin{equation}\label{M-LE}
\displaystyle \frac{\delta S}{\delta
\varphi^{\alpha\beta\gamma}}\equiv\square\varphi_{\alpha\beta\gamma}-\partial_{\alpha}\partial^\lambda\varphi_{\lambda\beta\gamma}-\partial_{\beta}\partial^\lambda\varphi_{\lambda\alpha\gamma}-\partial_{\gamma}\partial^\lambda\varphi_{\lambda\alpha\beta}=0\,.
\end{equation}
Taking the divergence of the equations, we come to the relations
\begin{equation}\label{M-LGI}
\displaystyle -\,\partial^\gamma\frac{\delta S}{\delta
\varphi^{\alpha\beta\gamma}}-\frac{1}{2}(\partial_\alpha\tau_\beta+\partial_\beta\tau_\alpha)\equiv0\,, \qquad
\tau_\alpha\equiv2\,\partial^\beta\partial^\gamma\varphi_{\alpha\beta\gamma}\, .
\end{equation}
These relations mean that vector field $\tau_\alpha$ obeys Killing
equations on shell. The space of Killing vectors is ten-dimensional
for $d=4$ Minkowski space. Once the linear combination of the
differential consequences of Lagrangian equations reduces to the
action of the operators with finite kernel onto the functions of
fields and  their derivatives, relations (\ref{M-LGI}) represent the
modified gauge identities (\ref{GI}) of the model (\ref{M-LS}).

Given the modified gauge identity (\ref{GI}), one can find the
unfree gauge symmetry transformations by recipe (\ref{Unfree}),
(\ref{GPE}). For the action (\ref{M-LS}), this recipe leads to the
transformations
\begin{equation}\label{M-LGS}
\delta_\epsilon\varphi_{\alpha\beta\gamma}=\frac{1}{3}\big(\partial_\alpha\epsilon_{\beta\gamma}
+ \partial_\beta\epsilon_{\gamma\alpha}
+\partial_\gamma\epsilon_{\alpha\beta}\big)\,,
\end{equation}
where $\epsilon_{\alpha\beta}$ is symmetric tracefull tensor obeying
transversality equations
\begin{equation}\label{M-LGPE}
    \partial_\beta\epsilon^{\beta\alpha}=0 \, .
\end{equation}
So, one can see that proceeding from the modified gauge identities
(\ref{M-LGI}) and following the general prescription of this
section, one arrives to the transformations (\ref{M-LGS}),
(\ref{M-LGPE}) found in the original work \cite{Campoleoni2013}. The
reducible counterpart of this symmetry can be also derived following
the prescriptions of this section. As the derivation is a bit
tedious, we do not include that. The reducible gauge symmetry
transformations include the third order derivatives of the gauge
parameter being the fourth-rank tensor with the Young tableau of the
window type. These transformations have been found in the article
\cite{FRANCIA2014248} by slightly different method.

Let us discuss the modular parameters and global conserved
quantities being analogous of the cosmological constant in the case
of the theory (\ref{M-LS}).  These conserved quantities have been
previously unknown. Relations (\ref{M-LGI}) identify
$\tau_{\alpha}=2\,\partial^\beta\partial^\gamma\varphi_{\alpha\beta\gamma}$ as the
completion function, which satisfies on shell Killing equations.
According to the general recipe (\ref{tau}), we have the on-shell
trivial quantities:
\begin{equation}\label{M-Ltau}
    \mathcal{T}_{\alpha}\equiv\tau_{\alpha} - u_{\alpha}(x,\Lambda )\approx
    0\, ,
\end{equation}
where $u_{\alpha}$ is a general solution to the Killing equations
in Minkowski space,
\begin{equation}\label{M-LKilling}
\displaystyle \partial_\alpha u_\beta+\partial_\beta u_\alpha=0 \quad
\Rightarrow \quad u_\alpha=\Lambda_\alpha+\Lambda_{\alpha\beta}x^{\beta}\,, \quad
\Lambda_{\alpha\beta}=-\Lambda_{\beta\alpha}\, ,
\end{equation}
and $\Lambda_\alpha$, $\Lambda_{\alpha\beta}$ are arbitrary constants.
According to the general recipe (\ref{J}), the constants $\Lambda$
should be expressed from relations (\ref{M-LKilling}) and their
differential consequences in terms of the fields and their
derivatives. Complementing relations (\ref{M-Ltau}) by the first
order consequences
\begin{equation}\label{M-Ltau1}
    \partial_\alpha \mathcal{T}_\beta-\partial_\beta \mathcal{T}_\alpha\approx
    0\, ,
\end{equation}
and substituting (\ref{M-LKilling}) into (\ref{M-Ltau}),
(\ref{M-Ltau1}), one can find all the constants $\Lambda$:
\begin{equation}\label{M-LJLambda}
    \Lambda_\alpha\approx J_\alpha\,,\quad \Lambda_{\alpha\beta}\approx J_{\alpha\beta}\,,
\end{equation}
where $J$ read
\begin{equation}\label{M-LJ1}
\displaystyle  J_\alpha\equiv
2\partial^\beta\partial^\gamma\varphi_{\alpha\beta\gamma}+
\big(\partial_\alpha\partial^\gamma\partial^\lambda\varphi_{\beta\gamma\lambda}-
\partial_\beta\partial^\gamma\partial^\lambda\varphi_{\alpha\gamma\lambda}\big)x^\beta\, ,\end{equation}
\begin{equation}\label{M-LJ2}
J_{\alpha\beta}\equiv-\big(\partial_\alpha\partial^\gamma\partial^\lambda
\varphi_{\beta\gamma\lambda}-\partial_\beta\partial^\gamma\partial^\lambda
\varphi_{\alpha\gamma\lambda}\big) \, .
\end{equation}
As a consequence of the unfree gauge symmetry, the theory
(\ref{M-LS}) in $d=4$ admits ten specific conserved quantities
$J_{\alpha}, J_{\alpha\beta}$. These quantities are not conserved currents, or
charges of any conserved current --- they represent a different type
of conserved quantity. Being defined at every point $x$ of
space-time, they remain on-shell constants in the sense that they do
not change from point to point on shell, much like the cosmological
constant of unimodular gravity. An interesting feature of the
conserved quantity (\ref{M-LJ1}) is the explicit $x$-dependence.
This is a general feature of the theories with unfree gauge symmetry
which admit tertiary and higher order constraints in Hamiltonian
formalism \cite{Abakumova:2020ajc}. In this article, we restrict
consideration of reducible constraints in Hamiltonian formalism by
the case without tertiary constraints. The model (\ref{M-LS}) gives
rise to the tertiary constraints, so the Hamiltonian analysis will
be done elsewhere.

Finalizing  this example, let us comment on the global conserved
quantities in the similar models to (\ref{M-LS}) for the tensor
fields of higher ranks proposed in Ref. \cite{Campoleoni2013}. The
model for rank $r$ tracefull symmetric tensor admits the completion
function being  double divergence of the field. This rank $r-2$
tensor on shell obeys Killing tensor equations. In Minkowski space,
any  Killing tensor decomposes into the product of Killing vectors.
Therefore, there exists $10\times(r-2)$ global conserved quantities.

\section{Unfree and reducible gauge symmetry in the constrained Hamiltonian formalism}\label{Section3}

In this section, we first briefly outline the general scheme of
deriving the unfree gauge symmetry for Hamiltonian constrained
systems. After that, we derive the higher order  reducible gauge
symmetry for the same constrained system without any restrictions
imposed on the gauge parameters. Then we establish connection
between these two symmetries.

The action of constrained Hamiltonian system reads
\begin{equation}\label{SH}
S[q(t),p(t), \lambda (t)] = \int dt \big( p_i \dot{q}^i
-H_T(q,p,\lambda)\big)\,, \qquad H_T(q,p,\lambda)= H(q,p)+
T_\alpha(q,p)\,\lambda^\alpha \, ,
\end{equation}
where the time-dependence is made explicit, while dependence on
space points is implicit. Summation over any condensed index
includes integration over space. The action and the Hamiltonian $H$
are supposed integrated over the space.

Assume that requirement of stability of the primary constraints has
consequences
\begin{equation}\label{Inv1}
\{\,T_\alpha(q,p), H_T(q,p,\lambda)\}
=T_\beta(q,p)\,W_1{}^\beta{}_\alpha(q,p,\lambda)
\,+\,\tau_a(q,p)\,\Gamma_1{}^a{}_\alpha(q,p,\lambda)\,,
\end{equation}
where $\Gamma_1{}^a{}_\alpha(q,p,\lambda)$ is the differential
operator with finite-dimensional kernel. These relations should not define any Lagrange multiplier, otherwise we would have the second class constraints. The second class case is not considered in this article.  Hence, to make the primary
constraints consistent with Hamiltonian equations, one has to assume
\begin{equation}\label{Htau}
\tau_a(q,p)- \Lambda_\mathcal{A} u^\mathcal{A}{}_a \approx 0 \, ,
\end{equation}
where $\Lambda_\mathcal{A}$ are the constant parameters, and $u^\mathcal{A}{}_a$ are the
elements of the generating set for the kernel of
$\Gamma_1{}^a{}_\alpha(q,p,\lambda)$, cf. (\ref{tau}). Relations
(\ref{Htau}) represent the secondary constraints, with one subtlety:
they involve arbitrary constants $\Lambda_\mathcal{A}$ which do not contribute
to the involution relations (\ref{Inv1}). In fact, $\tau_a$ are
defined by relations (\ref{Inv1}) modulo the kernel of
$\Gamma_1{}^a{}_\alpha(q,p,\lambda)$. So, we redefine $\tau_a$
absorbing the term $\Lambda_\mathcal{A} u^\mathcal{A}{}_a$. Hence, the secondary constraints are
assumed $\Lambda$-dependent. Once we have got the secondary
constraints $\tau_a(q,p,\Lambda)\approx 0$, they should conserve on
shell. Let us assume that no tertiary constraints appear, while  the
Lagrange multipliers remain undefined. This means the following
involution relations for the secondary constraints:
\begin{equation}\label{Inv2}
\{\,\tau_a(q,p), H_T(q,p,\lambda)\}
=T_\alpha(q,p)\,W_2{}^\alpha{}_a(q,p,\lambda)
\,+\,\tau_b(q,p)\,\Gamma_2{}^b{}_a(q,p,\lambda) \, .
\end{equation}
The $\Lambda$-dependent secondary constraints $\tau_a(q,p,\Lambda)$ are understood as Hamiltonian form of the completion functions described in the previous section.  Termination of the Dirac-Bergmann algorithm (\ref{Inv2}) means the modified Noether identity (\ref{GI}) between the Hamiltonian equations and primary constraints, being the variational derivatives of the Hamiltonian action (\ref{SH}), and the secondary constraints $\tau_a(q,p)$. The latter ones play the role of completion functions because they cannot be expressed as differential consequences of the variational equations, while $\tau$'s vanish on shell. This fact results from the assumption that operator $\Gamma_1{}^a{}_\alpha$ (\ref{Inv1}) has a finite-dimensional kernel. Given the identity, following the general recipe (\ref{Unfree}), (\ref{GPE}), we arrive to the unfree gauge symmetry transformations of the Hamiltonian action (\ref{SH}):
\begin{equation}\label{Hunfree1}
\varphi^i =(q,p)\qquad
\delta_\epsilon \varphi^i =
\{\varphi^i,T_\alpha \}\,\epsilon^\alpha\,+\, \{\varphi^i,\tau_a\}\,\epsilon{\,}^a\,,
\end{equation}
\begin{equation}
\label{Hunfree2}
\delta_\epsilon \lambda^\alpha=
\dot{\epsilon}^\alpha +
W_1{}^\alpha{}_\beta (q,p,\lambda)\,\epsilon^\beta
+W_2{}^\alpha{}_a(q,p,\lambda)\,\epsilon{\,}^a\,,
\end{equation}
\begin{equation}
\label{HGPE}
\dot{\epsilon}{\,}^a
\,+\,\Gamma_2{}^a{}_b(q,p,\lambda)\,\epsilon{\,}^b\,
\,+\,\Gamma_1{}^a{}_\alpha(q,p,\lambda)\,\epsilon^\alpha
\, =0\,.
\end{equation}
One can see, that both primary constraints $T_\alpha$ and the secondary ones $\tau_a$ generate gauge transformations with different gauge parameters. These parameters, however, are subject to equations (\ref{HGPE}) that relate them to each other. Given the gauge parameters $\epsilon^\alpha$ of the transformations generated by the primary constraints $T_\alpha$, the Cauchy problem is well-defined for the parameters $\epsilon^a$ of the transformations generated by the secondary constraints $\tau_a$. From this perspective, the gauge parameters $\epsilon^a$ are defined by $\epsilon^\alpha$ modulo initial data.
As demonstrated in the article \cite{Abakumova:2019uoo},
the gauge transformations (\ref{Hunfree1}), (\ref{Hunfree2}) indeed leave the Hamiltonian action unchanged
iff the gauge parameters obey eqs. (\ref{HGPE}).

Once relations (\ref{HGPE}) define the parameters $\epsilon^a$,
given $\epsilon^\alpha$, the admissible gauge conditions
$\chi^\alpha$ should fix only $\epsilon^\alpha$. Hence, the gauge
\begin{equation}\label{chi}
\dot{\lambda}{}^\alpha-\chi^\alpha(\varphi)=0
\end{equation}
is non-degenerate if the rectangular matrix $\big(\{\chi^\alpha,
T_\beta\}| \{\chi^\alpha, \tau_a\}\big)$ has the maximal rank.

The right hand sides of involution relations (\ref{Inv1}), (\ref{Inv2}) are at most linear in $\lambda$, so the structure coefficients involved in the gauge transformations (\ref{Hunfree1}), (\ref{Hunfree2}), (\ref{HGPE}) can be expanded w.r.t. the Lagrange multipliers:
\begin{equation}\label{W1}
W_1{}^\beta{}_\alpha{}
=V^\beta{}_\alpha\,-\,U^\beta{}_{\gamma\alpha}\,\lambda^\gamma\,,
\qquad\qquad
\Gamma_1{}^a{}_\alpha = V^a{}_\alpha
\,-\,U^a{}_{\beta\alpha}\,\lambda^\beta\,,
\end{equation}
\begin{equation}\label{W2}
W_2{}^\alpha{}_a=V^\alpha{}_a{}
\,-\,U^\alpha{}_{\beta a}\,\lambda^\beta\,,
\qquad\qquad
\Gamma_2{}^b{}_a{}=V^b{}_a{}
\,-\,U^b{}_{\alpha a}\,\lambda^\alpha\,.
\end{equation}
Substituting the expansion into (\ref{Inv1}), (\ref{Inv2}), one arrives to the involution relations
\begin{eqnarray}\label{irinv1}
\{T_\alpha,H\} &=& T_\beta\,V^\beta{}_\alpha(q,p)\,+\,\tau_a\,V^a{}_\alpha(q,p)\,, \\\label{irinv2}
\{\tau_a,H\} &=& T_\alpha\,V^\alpha{}_a(q,p)\,+\,\tau_b\,V^b{}_a(q,p)\,,\\\label{irinv3}
\{T_\alpha,T_\beta\} &=& T_\gamma\,U^\gamma{}_{\alpha\beta}(q,p)
\,+\,\tau_a\,U^a{}_{\alpha\beta}(q,p)\,,\\\label{irinv4}
\{T_\alpha,\tau_a\} &=& T_\beta\, U^\beta{}_{\alpha a}(q,p)
\,+\,\tau_b\,U^b{}_{\alpha a}(q,p)\,,\\\label{irinv5}
\{\tau_a,\tau_b\} &=& T_\alpha\,U^\alpha{}_{ab}(q,p)\,+\,\tau_c\,U^c{}_{ab}(q,p)\,.
\end{eqnarray}
These relations correspond to the unfree gauge symmetry in the case when $\Gamma_1{}^a{}_\alpha$  (\ref{W1}) is a differential operator with finite kernel.  It is the coefficient at secondary constraint $\tau_a$ in the stability conditions of the primary constraints (\ref{Inv1}) which is the source of the restrictions imposed on the gauge parameters. This operator does not admit local inverse. Therefore, the secondary constraints  \emph{ are not differential} consequences of the primary ones, even though they vanish on shell. This means, the system admits completion functions, and hence, the gauge symmetry is unfree.

As explained in Section \ref{Section2}, any variational system with unfree gauge symmetry always admits the  gauge symmetry with unconstrained gauge parameters, though with higher order derivatives, see relations (\ref{rho}), (\ref{GSR}). In the Hamiltonian setup, the general recipe (\ref{rho}), (\ref{GSR}) of constructing the gauge symmetry with unconstrained parameters can be implemented in a uniform way. This is done by absorbing the $\lambda$-independent part of non-invertible differential operator $\Gamma_1{}^a{}_\alpha$  (\ref{W1}) into the secondary  constraints. So, we introduce another set of the secondary constraints
\begin{equation}\label{tildetau}
\widetilde{\tau}_\alpha \equiv \tau_a(q,p) \,V^a{}_\alpha(q,p)\,.
\end{equation}
The constraints $\widetilde{\tau}_\alpha (q,p)\approx 0$ are the \emph{differential} consequences of the primary constraints $T_\alpha\approx 0$, unlike the irreducible secondary constraints $\tau_a$. This is because  $V^a{}_\alpha$ (\ref{tildetau}) does not admit, in general, the inverse in the class of differential operators.
Let us detail the distinction between $\widetilde{\tau}_\alpha$ and $\tau_a$.  If the secondary constraints $\tau_a$ were the differential consequences of the variational equations for the action (\ref{SH}), this would mean the differential operator $\psi^\alpha{}_a$ exists such that
\begin{equation}\label{inverseV}
  V^b{}_\alpha\psi^\alpha{}_a=\delta^b_a \, .
\end{equation}
For the case of the unfree gauge symmetry, this is impossible as $V^b{}_\alpha $ is assumed to have finite-dimensional right kernel.

Now, we make simplifying assumption that the primary constraints $T_\alpha$ are involutive by themselves. This means, the involution relations of the primary constraints (\ref{irinv3}) do not include secondary ones on the right hand side. In this setup, conservation condition of the primary constraints reads
\begin{equation}\label{HTtildetau}
\{\,T_\alpha(q,p), H_T(q,p,\lambda)\}
= \widetilde{\tau}_\alpha(q,p) \,+\, T_\beta(q,p)\,{W}_1{}^\beta{}_\alpha(q,p,\lambda)\,.
\end{equation}
Once the constraints $\tau_a$ are conserved, the constraints $\widetilde{\tau}_\alpha$, being the consequences of $\tau_a$ (see (\ref{tildetau})), are conserved as well
\begin{equation}\label{Htildetau2}
\{\,\widetilde{\tau}_\alpha(q,p), H_T(q,p,\lambda)\}
=T_\beta(q,p)\,\widetilde{W}_2{}^\beta{}_\alpha(q,p,\lambda)
\,+\,\widetilde{\tau}_\beta(q,p)\,\widetilde{\Gamma}_2{}^\beta{}_\alpha(q,p,\lambda)\,,
\end{equation}
where the structure functions of the algebra of constraints $\widetilde{\tau}_\alpha$ (\ref{tildetau}) are connected with the structure functions of the algebra (\ref{irinv1}), (\ref{irinv2}), (\ref{irinv4}) of the independent constraints $\tau_a$ by the relations
\begin{equation}\label{HGammatilde}
V^a{}_\beta\,\widetilde{\Gamma}_2{}^\beta{}_\alpha =
\{V^a{}_\alpha,H_T\}
+\Gamma_2{}^a{}_b\,V^b{}_\alpha
-T_\beta\,K{}^{\beta a}{}_\alpha
+  \tau_b\,  K{}^{[ab]}{}_\alpha\,;
\end{equation}
\begin{equation}\label{Wtilde}
\widetilde{W}_2{}^\alpha{}_\beta
=
W_2{}^\alpha{}_a\,V^a{}_\beta
+T_\gamma\,K{}^{[\alpha \gamma]}{}_\beta
+\tau_a\,K{}^{\alpha a}{}_\beta\,.
\end{equation}

Let us discuss now the reducibility of the secondary constraints  $\widetilde{\tau}_\alpha$ (\ref{tildetau}).
The structure coefficients $V^a{}_\alpha$ (\ref{irinv1}) connect the secondary irreducible constraints $\tau_a$ (the latter are not differential consequences of the primary constraints, as $V^a{}_\alpha$ have no local inverse operators) with another set of secondary constraints, $\widetilde{\tau}_\alpha$ (\ref{tildetau}), can be reducible on shell. This means, $\exists\, Z_1{}^\alpha{}_A$:
\begin{equation}\label{KERV}
V^a{}_\alpha\,Z_1{}^\alpha{}_A =
T_\alpha\,\kappa{}^{\alpha a}{}_A + \tau_b\,\kappa{}^{[ab]}{}_A \, .
\end{equation}
The null-vectors $Z_1{}^\alpha{}_A$ constitute the generating set of the left kernel of the structure coefficients $V^a{}_\alpha$.
Given the null-vectors, we see that corresponding linear combinations of the secondary constraints $\widetilde{\tau}_\alpha$ vanish modulo primary ones,
\begin{equation}\label{Z}
\widetilde{\tau}_\alpha\,Z_1{}^\alpha{}_A=T_\alpha\,\zeta{}^\alpha{}_A \, ,
\end{equation}
where $\zeta{}^\alpha{}_A$ are on shell trivial coefficients,
\begin{equation}\label{zeta}
\zeta{}^\alpha{}_A =T_\beta\,\kappa{}^{[\alpha \beta]}{}_A
+\tau_a\,\kappa{}^{\alpha a}{}_A\,,
\end{equation}
and the structure functions $ \kappa{}^{\alpha a}{}_A$ are defined by relations (\ref{KERV}).
The identities (\ref{Z}) conserve in time. This results in the consequences
\begin{equation}\label{Gammatilde}
\{Z_1{}^\alpha{}_A, H_T\}\approx
\zeta{}^\alpha{}_A
-\widetilde{\Gamma}_2{}^\alpha{}_\beta\,Z_1{}^\beta{}_A
-Z_1{}^\alpha{}_B\,\widetilde{\Gamma}^B{}_A\,.
\end{equation}
The identities (\ref{Z}) can be further reducible
\begin{equation}\label{Z1}
\begin{array}{l}
Z_1{}^\alpha{}_A\,Z_2{}^A{}_{A_1}
=T_\beta\,
U^{(1)}{}^{\beta\alpha}{}_{A_1}
+\widetilde{\tau}_\beta\,
U^{(2)}{}^{[\beta\alpha]}{}_{A_1} \, ;
\\[2mm]
\zeta{}^\alpha{}_A\,Z_2{}^A{}_{A_1}=
T_\beta\,U^{(3)}{}^{[\alpha\beta]}{\!}_{A_1}
+\widetilde{\tau}_\beta\,
U^{(1)}{}^{\alpha\beta}{}_{A_1} \,.
\end{array}
\end{equation}
Given the involution relations (\ref{HTtildetau}),  (\ref{Htildetau2}), and identities between primary and secondary constraints (\ref{Z}), certain combinations of the first and second time derivatives of the primary constraints identically reduce to the combinations of Hamiltonian equations and primary constraints. This fact means the Noether identities for the Hamiltonian action (\ref{SH}). Explicitly, the identities read
\begin{equation}\label{HredNI1}
Z_1{}^\alpha{}_A \{\varphi^i , T_\alpha  \}
\frac{\delta S}{\delta \varphi^i} -
\bigg[Z_1{}^\alpha{}_A \frac{d}{dt}
-Z_1{}^\beta{}_A {W}_1{}^\alpha{}_\beta
-\zeta^\alpha{}_A\bigg]  \frac{\delta S}{\delta \lambda^\alpha}\, \equiv 0 \, ;
\end{equation}
\begin{equation}\label{RedNI2}
 \bigg[\bigg( \delta^\beta_\alpha\,\frac{d}{dt} - \widetilde{\Gamma}_2{}^\beta{}_\alpha\bigg)\{\varphi^i , T_\beta \}
+ \{\varphi^i,\widetilde{\tau}_\alpha\}\bigg]\frac{\delta S}{\delta \varphi^i}
-
\bigg[\bigg( \delta^\beta_\alpha\,\frac{d}{dt} - \widetilde{\Gamma}_2{}^\beta{}_\alpha\bigg)
\bigg( \delta^\gamma_\beta\,\frac{d}{dt}
 -  \,{W}_1{}^\gamma{}_\beta \bigg)
-\widetilde{W}_2{}^\gamma{}_\alpha\bigg]
 \frac{\delta S}{\delta \lambda^\gamma}
 \equiv 0 \, .
\end{equation}

By virtue of the second Noether theorem, the identities between variational equations are equivalent to the gauge symmetry of the action. Contracting the above identities with arbitrary gauge parameters, and integrating by parts to switch the derivatives from the equations, one can find the gauge variations of the variables, being the coefficients at corresponding variational equations. In this way, proceeding from the above identities, we arrive to the gauge transformations of the phase-space variables $\varphi^i$ and Lagrange multipliers $\lambda^\alpha$:
\begin{equation}\label{RedGS1}
\delta_\varepsilon \varphi^i\,=\,
\{\varphi^i , T_\alpha  \}\,
\bigg(
\dot{\varepsilon}{}^\alpha
\,+\,\widetilde{\Gamma}_2{}^\alpha{}_\beta \, \varepsilon^\beta
-Z_1{}^\alpha{}_A\,\varepsilon^A
\bigg)
\,-\,  \{\varphi^i, \widetilde{\tau}_\alpha \}\, \varepsilon^\alpha\,;
\end{equation}
\begin{equation}\label{RedGS2}
\delta_\varepsilon \lambda^\alpha\,=\,
\bigg(
\delta^\alpha_\beta\frac{d}{dt}
\,+\, \,{W}_1{}^\alpha{}_\beta
\bigg)\,
\bigg( \dot{\varepsilon}{}^\beta
\,+\, \,\widetilde{\Gamma}_2{}^\beta{}_\gamma\,\varepsilon^\gamma
-Z_1{}^\beta{}_A\,\varepsilon^A
\bigg)
\,-\, \widetilde{W}_2{}^\alpha{}_\beta
\,\varepsilon^\beta
\,-\,\zeta{}^\alpha{}_A\,\varepsilon^A \, .
\end{equation}
Now, let us explain the reducibility of the  Noether identities (\ref{HredNI1}), (\ref{RedNI2}), and hence the symmetry of symmetry of the gauge transformations above.
Notice, that the identity (\ref{HredNI1}) originates from the reducibility conditions of the secondary constraints (\ref{Z}). The secondary constraints $\widetilde{\tau}_\alpha$ are the first order differential consequences of the primary constraints and Hamiltonian equations. These are the variational equations. Expressing relations (\ref{Z}) in terms of variational equations of the Hamiltonian action (\ref{SH}) and their time derivatives, one arrives at the Noether identity (\ref{HredNI1}). This first order identity results in the gauge symmetry transformations with the first order derivatives of the gauge parameters $\varepsilon^A$. The second group of Noether identities (\ref{RedNI2}) originates from the fact that the time derivatives of the secondary constraints, being the second order consequences of the variational equations, reduce to the combinations of secondary and primary constraints. This amounts to saying that involution relations (\ref{HTtildetau}), (\ref{Htildetau2}) result in the second order Noether identities (\ref{RedNI2}). These identities lead to the gauge symmetry transformations with the second order time derivatives of the gauge parameters $\varepsilon^\alpha$.
Taking the time derivative of the reducibility relations (\ref{Z}), one comes to the identities involving the time derivatives of the secondary constraints $\widetilde{\tau}_\alpha$.
On the other hand, the identities (\ref{RedNI2}) express the fact that the time derivatives of $\widetilde{\tau}_\alpha$ reduce to the  equations of motion and their first order consequences. Hence, the identities (\ref{HredNI1}) and (\ref{HredNI1}) are reducible, and therefore corresponding gauge symmetry transformations should have the symmetry of symmetry. The parameters of this symmetry of symmetry will be denoted $\omega^A$. The null-vectors $Z_1{}^\alpha{}_A$ of the secondary constraints serve as the leading terms of this identity of identities, and hence contribute to the symmetry of symmetry. The same null-vectors $Z_1{}^\alpha{}_A$ simultaneously  serve as the leading terms of generators of the identity (\ref{HredNI1}), originating from the relations (\ref{Z}). Once the null-vectors $Z_1{}^\alpha{}_A$ are reducible (\ref{Z1}), the identities (\ref{HredNI1}), being the variational form of relations (\ref{Z}) are reducible as such, irrespectively to (\ref{RedNI2}). This reducibility of the identities leads to the symmetry of symmetry transformations with parameters $\omega^{A_1}$. As a result, we arrive to the following gauge symmetry transformations of the original gauge parameters
\begin{equation}\label{SofS1}
\delta_\omega \varepsilon^\alpha = Z_1{}^\alpha{}_B\,\omega^B\,,
\qquad
\delta_\omega \varepsilon^A =
\dot{\omega}^A
-\widetilde{\Gamma}^A{}_B\,\omega^B
-Z_2{}^A{}_{A_1}\,\omega^{A_1}\,,
\end{equation}
where the structure functions $\widetilde{\Gamma}^A{}_B$  are defined by relations (\ref{Gammatilde}).
These symmetries are reducible in their own turn due to identities (\ref{Z1}) between null-vectors $Z_1{}^\alpha{}_A$:
\begin{equation}\label{SofS2}
\delta_\eta \omega^A = Z_2{}^A{}_{A_1}\,\eta^{A_1}\,,
\qquad
\delta_\eta \omega^{A_1} =
\dot{\eta}^{A_1}
-\widetilde{\Gamma}^{A_1}{}_{B_1}\,\eta^{A_1}\,,
\end{equation}
where the structure functions $\widetilde{\Gamma}^{A_1}{}_{B_1}$  are defined by relations
\begin{equation}\nonumber
\{{Z}_2{}^A{}_{A_1},H_T\} \approx
\widetilde{\Gamma}^A{}_{B}\, Z_2{}^B{}_{A_1}
-Z_2{}^A{}_{B_1}\,\widetilde{\Gamma}^{B_1}{}_{A_1}\,.
\end{equation}
Obviously, the transformations of parameters $\omega^A$, $\omega^{A_1}$ do not affect on the transformations of the original gauge parameters $\varepsilon^\alpha$, $\varepsilon^A$,
\begin{equation}
\delta_\eta(\delta_\omega \varepsilon^\alpha) \approx 0 \, ,
\qquad
\delta_\eta(\delta_\omega \varepsilon^A) \approx 0 \, .
\end{equation}
As we have seen above, the same action of constrained Hamiltonian system (\ref{SH}) can admit two different descriptions of gauge symmetry: (i) the unfree gauge transformations (\ref{Hunfree1}), (\ref{Hunfree2}) with the first order time derivatives of the gauge parameters restricted by the differential equations (\ref{HGPE}) of the first order with respect to time, and (ii) the reducible gauge symmetry (\ref{RedGS1}), (\ref{RedGS2}) with higher order time derivatives of unrestricted gauge parameters. The source of the unfree gauge symmetry in Hamiltonian formalism is in certain specifics of the Dirac-Bergmann algorithm. If the secondary constraints $\tau_a$ arise in the stability conditions of the primary ones with the coefficient being the differential operator with finite-dimensional kernel ($\Gamma^a{}_\alpha$ of relations (\ref{Inv1})), then the gauge symmetry generated by the constraints turns out unfree, see (\ref{Hunfree1}), (\ref{Hunfree2}),  (\ref{HGPE}). One more consequence of the fact that $\Gamma^a{}_\alpha$ does not admit local inverse is that the secondary constraints $\tau_a$ are forced to coincide with certain element of the kernel of $\Gamma^a{}_\alpha$. Hence, the secondary constraints $\tau_a$ depend on the modular parameters $\Lambda$ labelling the kernel elements. The alternative form of the gauge symmetry implies that the $\lambda$-independent term $V^a{}_\alpha$ of the irreversible operator $\Gamma^a{}_\alpha$ is absorbed by the secondary constraints (see (\ref{tildetau})). These alternative secondary constraints $\widetilde{\tau}_\alpha$ can be reducible, while  the $\Lambda$-dependent constraints $\tau_a$ are irreducible.
Unlike the irreducible constraints, the reducible ones are the differential consequences of the primary constraints, see (\ref{HTtildetau}). Hence, the second Noether theorem applies to the case when $\widetilde{\tau}_\alpha$ are treated as secondary constraints. By virtue of the theorem, one arrives at the unconstrained gauge symmetry, though with the second order derivatives of gauge parameters (\ref{RedGS1}), (\ref{RedGS2}), and reducibility gauge transformations (\ref{SofS1}), (\ref{SofS2}).

The symmetries with the higher order derivatives of gauge parameters are long known in the constrained Hamiltonian formalism \cite{Anderson1951B},\cite{Utiyama1959},\cite{Mukunda1980},\cite{Castellani1982}. The reducible higher order gauge symmetries have been never considered in Hamiltonian framework, to the best of our knowledge. As we have seen, the unfree gauge symmetry (\ref{Hunfree1}), (\ref{Hunfree2}), (\ref{HGPE}) can be always converted into the reducible gauge symmetry with the higher order time derivatives of unconstrained gauge parameters. This gauge symmetry has some specifics comparing to the usual first order reducible gauge symmetry. In particular, the null-vectors $Z_1{}^\alpha{}_A$ of the secondary constraints (\ref{Z}) contribute to the reducible gauge symmetry twice. First, it generates the gauge transformations of the phase-space variables (\ref{RedGS1}), and also it contributes to the transformations of Lagrange multipliers with the second order time derivatives of the gauge parameters (\ref{RedGS2}), while in the case of reducible primary constraints the null-vectors are not involved in the transformations of the original phase space, and no second order time derivatives are involved of the gauge parameters. Second, the same null-vectors generate the gauge symmetry of gauge symmetry (\ref{SofS1}).
Similar specifics is observed again for the next level null-vectors $Z_2{}^{A}{}_{A_1}$, see (\ref{SofS1}), (\ref{SofS2}).
This specifics leads to corresponding changes in the sector of extra-ghosts in the Hamiltonian BFV-BRST formalism comparing to the usual case of reducible primary constraints, see in the next section.

As we have seen in this section at the level of constrained Hamiltonian formalism, any theory with unfree gauge symmetry always admits reducible form of gauge symmetry with higher order derivatives of unconstrained gauge parameters.
Let us discuss now connection between these two forms of gauge symmetry.
Consider unfree gauge variation (\ref{Hunfree1}), (\ref{Hunfree2}) of Hamiltonian action (\ref{SH})
\begin{equation}\label{HUnfreevariation}
\begin{array}{c}
\displaystyle \delta_\epsilon S =\int dt\, \bigg\{
\frac{ \delta S}{ \delta \varphi^i}\,\delta_\epsilon \varphi^i   +
 \frac{ \delta S}{ \delta \lambda^\alpha}\, \delta_\epsilon \lambda^\alpha   \bigg\}
\\[2mm]
\displaystyle =\int dt
\bigg\{
\frac{\delta S}{\delta \varphi^i}\,
\big[\delta_\epsilon \varphi^i - \{\varphi^i,T_\alpha \}\,\epsilon^\alpha\,-\, \{\varphi^i,\tau_a\}\,\epsilon{\,}^a\big]\\[2mm]
\displaystyle+\,\frac{\delta S}{\delta \lambda^\alpha}\,
\big[\delta_\epsilon \lambda^\alpha - \dot{\epsilon}^\alpha -
W_1{}^\alpha{}_\beta\,\epsilon^\beta
-W_2{}^\alpha{}_a\,\epsilon{\,}^a\big]
+\tau_a\, \big[\dot{\epsilon}{\,}^a
\,+\,\Gamma_2{}^a{}_b\,\epsilon{\,}^b\,
\,+\,V{}^a{}_\alpha\,\epsilon^\alpha\big]\bigg\} \, .
\end{array}
\end{equation}
This variation reduces to the integral of a total derivative upon account of the involution relations (\ref{Inv1}), (\ref{Inv2}) provided for the gauge parameters $\epsilon^\alpha,\, \epsilon^a$ to obey the equations (\ref{HGPE}).
Now, let change the parameters $\epsilon^\alpha$, $\epsilon^a$ of the unfree gauge transformation by the parameters $\varepsilon^\alpha$, $\varepsilon^A$:
\begin{equation}\label{substtutionepsilon}
\epsilon^\alpha =
\dot{\varepsilon}^\alpha
+ \widetilde{\Gamma}_2{}^\alpha{}_\beta\,\varepsilon^\beta
-Z_1{}^\alpha{}_A\,\varepsilon^A \, ,
\qquad
\epsilon^a = -V{}^a{}_\alpha\,\varepsilon^\alpha \, .
\end{equation}
Upon substitution of this parametrization into the unfree gauge variation (\ref{HUnfreevariation}),  it reads (modulo total derivative in the integrand)
\begin{equation}\label{Hredgaugevariation}
\begin{array}{c}
\displaystyle \delta_{\epsilon(\varepsilon)} S =\delta_\varepsilon S
\\[2mm]
\displaystyle =\int dt
\bigg\{
\frac{\delta S}{\delta \varphi^i}\,
\bigg[
\delta_\varepsilon \varphi^i -
\{\varphi^i , T_\alpha  \}\,
\bigg(
\dot{\varepsilon}{}^\alpha
\,+\,\widetilde{\Gamma}_2{}^\alpha{}_\beta \, \varepsilon^\beta
-Z_1{}^\alpha{}_A\,\varepsilon^A
\bigg)
+  \{\varphi^i, \widetilde{\tau}_\alpha \}\, \varepsilon^\alpha
\bigg]\\[2mm]
\displaystyle +\,\frac{\delta S}{\delta \lambda^\alpha}\,
\bigg[
\delta_\varepsilon \lambda^\alpha -
\bigg(
\delta^\alpha_\beta\frac{d}{dt}
\,+\, \,{W}_1{}^\alpha{}_\beta
\bigg)\,
\bigg( \dot{\varepsilon}{}^\beta
\,+\, \,\widetilde{\Gamma}_2{}^\beta{}_\gamma\,\varepsilon^\gamma
-Z_1{}^\beta{}_A\,\varepsilon^A
\bigg)
+ \widetilde{W}_2{}^\alpha{}_\beta
\,\varepsilon^\beta
+\zeta{}^\alpha{}_A\,\varepsilon^A
\bigg]\bigg\}\,.
\end{array}
\end{equation}
This expression is a gauge variation of the original action (\ref{SH}) with respect to the reducible gauge transformations
(\ref{RedGS1}), (\ref{RedGS2}). On the account of the involution relations (\ref{HTtildetau}), (\ref{Htildetau2}), (\ref{Z}) it vanishes for any $\varepsilon^\alpha$, $\varepsilon^A$. In this way, we see that substitution (\ref{substtutionepsilon}) resolves, in a sense, the equations (\ref{HGPE}) imposed on the unfree gauge symmetry parameters. This substitution, as one can see from (\ref{Hredgaugevariation}), leads to the identically vanishing gauge variation without any restrictions on gauge parameters. This means, the unfree gauge transformations (\ref{Hunfree1}), (\ref{Hunfree2}) are converted by the substitution (\ref{substtutionepsilon}) into the reducible transformations (\ref{RedGS1}), (\ref{RedGS2}) modulo trivial ones (such that vanish on shell).

Notice, that substitution (\ref{substtutionepsilon}) absorbs the structure coefficient $V^a{}_\alpha$ of the involution relations (\ref{irinv1}) into the gauge parameters. This is dual to the absorption of the same coefficient by the reducible constraints $\widetilde{\tau}_\alpha$ (\ref{tildetau}). If the operators $V^a{}_\alpha$ were invertible (they are not as the null-vectors exist, see (\ref{KERV})), in the sense of relation (\ref{inverseV}), the substitution (\ref{substtutionepsilon}) would mean that the gauge parameters $\epsilon^\alpha$ and $\epsilon^a$ can be expressed in terms of  unrestricted parameters $\varepsilon^\alpha, \varepsilon^A$ and their time derivatives $\dot{\varepsilon}{}^\alpha$. Then, we would have equivalence between the unfree gauge symmetry with the first order derivatives of gauge parameters, and the gauge symmetry with the second order derivatives of unrestricted gauge parameters.
Given the irreversible structure coefficient $V^a{}_\alpha$, the map of the unfree gauge symmetry to the reducible higher order symmetry is not invertible. Corresponding map is further discussed in the next section at the level of Hamiltonian BFV-BRST formalism.

\section{Connection between Hamiltonian BRST complexes
of the reducible and unfree gauge symmetry}\label{Section4}

At first, briefly remind the generalities of Hamiltonian BFV-BRST formalism for unfree gauge theory, for more details see \cite{Abakumova:2019uoo}, \cite{Abakumova:2020ajc}.
After that, we explain the construction of the BFV-BRST formalism for the reducible constrained system that generates gauge transformations with higher derivatives of gauge parameters. The main distinctions from the standard BFV-BRST formalism for the reducible systems of primary constraints \cite{teitelboim1992quantization}, \cite{Batalin:1983pz} concern the sector of extra-ghosts.  The irreducible constrained system generating the unfree gauge symmetry and the reducible one are connected to the same Hamiltonian action (\ref{SH}), hence the system admits two BRST complexes.

Given the constrained Hamiltonian system (\ref{SH}) with the involution relations (\ref{irinv1})--(\ref{irinv5}), the canonical pairs of minimal sector ghosts are assigned to primary and secondary constraints, $T_\alpha$ and $\tau_a$:
\begin{equation}
\displaystyle \{C^\alpha,\overline{P}_\beta\}=\delta^\alpha_\beta\,, \quad \text{gh}\,C^\alpha=-\,\text{gh}\,\overline{P}_\alpha=1\,, \quad \varepsilon(C^\alpha)=\varepsilon(\overline{P}_\alpha)=1\,;
\end{equation}
\begin{equation}
\displaystyle \{{C}^a,\overline{{P}}_b\}=\delta^a_b\,, \quad \text{gh}\,{C}^a=-\,\text{gh}\,\overline{{P}}_a=1\,, \quad \varepsilon({C}^a)=\varepsilon(\overline{{P}}_a)=1\,.
\end{equation}
The BRST charge of the minimal sector begins with the constraints
multiplied by the ghosts, and the higher orders of ghosts are sought from the Poisson-nilpotency equation,
\begin{equation}\label{Qminir}
\displaystyle Q_{\text{min}}=T_\alpha C^\alpha+\tau_a {C}^a + \cdots\,, \qquad \{Q_{\text{min}},Q_{\text{min}}\}=0\,.
\end{equation}
BRST invariant Hamiltonian $\mathcal{H}$ begins with the original one, while the ghost terms are sought from the requirement that $\mathcal{H}$ to commute with $Q_{\text{min}}$,
\begin{equation}\label{mathHir}
\displaystyle \mathcal{H}=H+\cdots\,, \qquad \{Q_{\text{min}},\mathcal{H}\}=0\,.
\end{equation}
In the non-minimal sector, the anti-ghosts and Lagrange multipliers are assigned only to the primary constraints $T_\alpha$, while the secondary constraints $\tau_a$ do not imply introduction of any auxiliary variables:
\begin{equation}\label{irnonminghosts}
\displaystyle \{P^\alpha,\overline{C}_\beta\}=\delta^\alpha_\beta\,, \quad \text{gh}\,P^\alpha=-\,\text{gh}\,\overline{C}_\alpha=1\,, \quad \varepsilon(P^\alpha)=\varepsilon(\overline{C}_\alpha)=1\,;
\end{equation}
\begin{equation}\label{irnonminlambda}
\displaystyle \{\lambda^\alpha,\pi_\beta\}=\delta^\alpha_\beta\,, \quad \text{gh}\,\lambda^\alpha=-\,\text{gh}\,\pi_\alpha=0\,, \quad \varepsilon(\lambda^\alpha)=\varepsilon(\pi_\alpha)=0\,.
\end{equation}
Complete BRST charge reads
\begin{equation}\label{Qir-nonmin}
\displaystyle Q=Q_{\text{min}}+\pi_\alpha P^\alpha\,.
\end{equation}
Gauge-fixing Fermion includes  the Lagrange multipliers to the primary constraints, and corresponding gauges $\chi^\alpha$, while the secondary constraints do not imply any gauges, nor do they have any Lagrange multipliers,
\begin{equation}\label{irredPsi}
\displaystyle \Psi=\overline{C}_\alpha\chi^\alpha+\overline{P}_\alpha \lambda^\alpha\,.
\end{equation}
Given the BRST invariant Hamiltonian and gauge Fermion, unitarizing Hamiltonian is constructed by the usual recipe
\begin{equation}\label{HPsiir}
\displaystyle H_\Psi=\mathcal{H}+\{Q,\Psi\}\,.
\end{equation}
Once the secondary constraints $\tau_a$ explicitly depend on the
modular parameters $\Lambda$ (see (\ref{Htau})), the BRST charge
involves these constants even though the original action is
$\Lambda$-independent. BRST invariants assigned to these constraints
are BRST-exact, hence no observables can be associated to them.
Therefore, the global conserved quantities $J$ (\ref{J}), being the
derivatives of $\tau_a$, will be also BRST-trivial with respect to
the BRST operator (\ref{Qminir}). Hence, this BRST complex describes
the dynamics on the fixed level surface of these global conserved
quantities. In this framework, the questions cannot be posed about
the spectrum of $\Lambda$, or quantum transitions between the states
with different values of $\Lambda$. As we shall see, the complex
related to the reducible gauge symmetry treats these quantities
differently.

Consider now the construction of Hamiltonian BRST complex related to the reducible secondary constraints $\widetilde{\tau}_a$ (\ref{tildetau}). At the level of minimal sector, the construction follows the standard prescription of BFV-method  \cite{Batalin:1983pz}.
The canonical ghost pairs are assigned to the primary constraints $T_\alpha$,
\begin{equation}\label{redghosts1}
\displaystyle \{C^\alpha,\overline{P}_\beta\}=\delta^\alpha_\beta\,, \quad \text{gh}\,C^\alpha=-\,\text{gh}\,\overline{P}_\alpha=1\,, \quad \varepsilon(C^\alpha)=\varepsilon(\overline{P}_\alpha)=1\,,
\end{equation}
and to the secondary constraints $\widetilde{\tau}_\alpha$ (\ref{tildetau}):
\begin{equation}\label{redghosts2}
\displaystyle \{\mathcal{C}^\alpha,\overline{\mathcal{P}}_\beta\}=\delta^\alpha_\beta\,, \quad \text{gh}\,\mathcal{C}^\alpha=-\,\text{gh}\,\overline{\mathcal{P}}_\alpha=1\,, \quad \varepsilon(\mathcal{C}^\alpha)=\varepsilon(\overline{\mathcal{P}}_\alpha)=1\,.
\end{equation}
Given the reducibility of secondary constraints (\ref{Z}), the ghosts for ghosts are introduced,
\begin{equation}\label{gg1}
\displaystyle \{\mathcal{C}{}^A,\overline{\mathcal{P}}_{B}\}=\delta^A_B\,, \quad \text{gh}\,\mathcal{C}{}^A=-\,\text{gh}\,\overline{\mathcal{P}}_{A}=2\,, \quad \varepsilon(\mathcal{C}{}^A)=\varepsilon(\overline{\mathcal{P}}_{A})=0\,.
\end{equation}
Reducibility of the null-vectors (\ref{Z1}) implies introduction of the next level ghosts:
\begin{equation}\label{gg2}
\displaystyle \{\mathcal{C}_{}^{A_1},\overline{\mathcal{P}}_{B_1}\}=\delta^{A_1}_{B_1}\,, \quad \text{gh}\,\mathcal{C}{}^{A_1}=-\,\text{gh}\,\overline{\mathcal{P}}_{A_1}=3\,, \quad \varepsilon(\mathcal{C}{}^{A_1})=\varepsilon(\overline{\mathcal{P}}_{A_1})=1\,.
\end{equation}
The BRST charge is constructed in the usual way, beginning with the constraints and null-vectors with the ghost terms defined by the equation of Poisson-nilpotency:
\begin{equation}\label{Qminr}
{Q}_{\text{min}}=T_\alpha C^\alpha+\widetilde{\tau}_\alpha\mathcal{C}^\alpha+\overline{\mathcal{P}}_\alpha Z_1{}^\alpha{}_A\mathcal{C}{}^A+\overline{\mathcal{P}}_{A}Z_2{}^A{}_{A_1}\mathcal{C}{}^{A_1}+\cdots, \quad \{{Q}_{\text{min}},{Q}_{\text{min}}\}=0\,.
\end{equation}
The BRST invariant Hamiltonian $\mathcal{H}$ is iteratively constructed by solving the equation of that $\mathcal{H}$ commutes with ${Q}_{\text{min}}$ starting from the original Hamiltonian:
\begin{equation}\label{mathHr}
\displaystyle {\mathcal{H}}=H+\cdots\,,
\quad \{{Q}_{\text{min}},\mathcal{H}\}=0\,.
\end{equation}
Much like the Hamiltonian, any other phase-space function, being an invariant of the reducible gauge symmetry (\ref{RedGS1}), is assigned with the BRST invariant extension. This extension is unique modulo BRST exact terms.

Now, let us discuss the gauge fixing and the ghosts of non-minimal
sector. The standard BFV construction \cite{Batalin:1983pz} of the
non-minimal ghost sector corresponds to the case with reducible
primary first class constraints. In this case, the gauge
transformations of the Lagrange multipliers involve the first
derivatives of the gauge parameters, while the corresponding
constraints are reducible. The higher order reducible gauge symmetry
(\ref{RedGS1}), (\ref{RedGS2}), (\ref{SofS1}), (\ref{SofS2}) is of a
different structure. The primary constraints $T_\alpha$ are
irreducible, and the gauge transformations of Lagrange multipliers
(\ref{RedGS2}) include the second order derivatives of the gauge
parameters. These transformations are reducible (\ref{SofS1}), while
the corresponding constraints are not. In the sector of canonical
variables, the primary constraints generate transformations with the
first derivatives of gauge parameters (\ref{RedGS1}). The secondary
constraints $\widetilde{\tau}_\alpha$ are reducible, and they
generate the gauge with the same gauge parameters, as $T_\alpha$,
though without time derivatives of $\varepsilon^\alpha$.  Once the
primary and secondary constraints share the same gauge parameters
$\varepsilon^\alpha$, they share the same gauge-fixing conditions
$\chi^\alpha(\varphi)$, cf. (\ref{chi}). Hence, the same pair of
anti-ghosts is introduced for the $T_\alpha$ and
$\widetilde{\tau}_\alpha$, even though their ghosts are different.
Also notice, that the gauges $\chi^\alpha$ are assumed irreducible
 as no null-vectors exist for the primary constraints. Hence, no extraghosts are introduced in the sector of these gauges,
 unlike the canonical BFV-scheme \cite{Batalin:1983pz}. Irreducibility of $\chi^\alpha$ implies that the Lagrange multipliers to primary constraints $T_\alpha$ are irreducible and do not require extra-ghosts.
 The reducible gauge symmetry (\ref{SofS1}), induced by the null-vector $Z_1{}^\alpha{}_A$,
 leads to imposing reducible gauge conditions $\omega_1{}^A{}_\alpha\mathcal{C}^\alpha$
 on ghosts $\mathcal{C}^\alpha$ assigned to secondary constraints $\widetilde{\tau}_\alpha$.
 Hence, gauge-fixing conditions are imposed on associated anti-ghosts and Lagrange multipliers,
 and the corresponding extra-ghosts are introduced. The gauge symmetry (\ref{SofS2}),
 induced by $Z_2{}^A{}_{A_1}$, implies imposing gauge conditions $\omega_2{}^{A_1}{}_A\mathcal{C}^A$ on ghosts $\mathcal{C}^A$.
 As $Z_2{}^A{}_{A_1}$ does not have its own null-vectors, no gauge-fixing conditions and extra-ghosts are needed for associated
 anti-ghosts and Lagrangian multipliers, like in the standard formalism \cite{Batalin:1983pz} for the second stage of reducibility.

So, the distinction from the canonical procedure
\cite{Batalin:1983pz} is that primary constraints $T_\alpha$ of the
original action (\ref{SH}) are irreducible, as well as the gauges
$\chi^\alpha$. Therefore, corresponding anti-ghosts and Lagrangian
multipliers do not require gauge-fixing conditions and related
extra-ghosts. Given this feature, we introduce canonical pairs of
anti-ghosts and Lagrange multipliers assigned to the primary
constraints $T_\alpha$ and corresponding gauges (\ref{chi}) as if
there were no reducibilities of gauge symmetry:
\begin{equation}\
\displaystyle \{P^\alpha,\overline{C}_\beta\}=\delta^\alpha_\beta\,, \quad \text{gh}\,P^\alpha=-\,\text{gh}\,\overline{C}_\alpha=1\,, \quad \varepsilon(P^\alpha)=\varepsilon(\overline{C}_\alpha)=1\,;
\end{equation}
\begin{equation}
\displaystyle \{\lambda^\alpha,\pi_\beta\}=\delta^\alpha_\beta\,, \quad \text{gh}\,\lambda^\alpha=-\,\text{gh}\,\pi_\alpha=0\,, \quad \varepsilon(\lambda^\alpha)=\varepsilon(\pi_\alpha)=0\,.
\end{equation}
Once the gauge parameters $\varepsilon$ enjoy the gauge symmetry
 (\ref{SofS1}) generated by $Z_1{}^\alpha{}_A$, corresponding ghosts need gauge
 fixing, so  related extra-ghosts are introduced in this sector:
\begin{equation}\label{extrag1}
\displaystyle \{\mathcal{P}{}^A,\overline{\mathcal{C}}_{B}\}=\delta^A_B\,, \quad \text{gh}\,\mathcal{P}{}^A=-\,\text{gh}\,\overline{\mathcal{C}}_{A}=2\,, \quad \varepsilon(\mathcal{P}{}^A)=\varepsilon(\overline{\mathcal{C}}_{A})=0\,;
\end{equation}
\begin{equation}\label{etral1}
\displaystyle \{\lambda^A,\pi_B\}=\delta^A_B\,, \quad
\text{gh}\,\lambda^A=-\,\text{gh}\,\pi_A=1\,, \quad
\varepsilon(\lambda^A)=\varepsilon(\pi_A)=1\,.
\end{equation}
The gauge symmetry (\ref{SofS1}) enjoys the symmetry (\ref{SofS2})
generated by $Z_2{}^A{}_{A_1}$, hence corresponding ghosts need
gauge fixing. This implies introduction of the extra-ghosts related
to these gauges, and corresponding Lagrange multipliers:
\begin{equation}\label{extrag2}
\displaystyle
\{\mathcal{P}_{}^{A_1},\overline{\mathcal{C}}_{B_1}\}=\delta^{A_1}_{B_1}\,,
\quad
\text{gh}\,\mathcal{P}{}^{A_1}=-\,\text{gh}\,\overline{\mathcal{C}}_{A_1}=3\,,
\quad
\varepsilon(\mathcal{P}{}^{A_1})=\varepsilon(\overline{\mathcal{C}}_{A_1})=1\,;
\end{equation}
\begin{equation}\label{extral2}
\displaystyle \{\lambda^{A_1},\pi_{B_1}\}=\delta^{A_1}_{B_1}\,, \quad
\text{gh}\,\lambda^{A_1}=-\,\text{gh}\,\pi_{A_1}=2\,, \quad
\varepsilon(\lambda^{A_1})=\varepsilon(\pi_{A_1})=0\,.
\end{equation}
We also introduce extra-ghosts
\begin{equation}
\displaystyle \{\lambda^{(1')A_1},\pi^{(1')}{}_{B_1}\}=\delta^{A_1}_{B_1}\,, \quad \text{gh}\,\lambda^{(1')A_1}=-\,\text{gh}\,\pi^{(1')}{}_{A_1}=1\,, \quad \varepsilon(\lambda^{(1')A_1})=\varepsilon(\pi^{(1')}{}_{A_1})=1\,,
\end{equation}
related to the gauge conditions $\sigma_2{}^{A_1}{}_A\lambda^A$
on Lagrange multipliers $\lambda^A$, and extra-ghosts
\begin{equation}\label{P1'A1C1'B1}
\displaystyle \{\mathcal{P}^{(1')A_1},\overline{\mathcal{C}}{}^{(1')}{}_{B_1}\}=\delta^{A_1}_{B_1}\,, \quad \text{gh}\,\mathcal{P}^{(1')A_1}=-\,\text{gh}\,\overline{\mathcal{C}}{}^{(1')}{}_{A_1}=2\,, \quad \varepsilon(\mathcal{P}^{(1')A_1})=\varepsilon(\overline{\mathcal{C}}{}^{(1')}{}_{A_1})=0\,,
\end{equation}
to gauge conditions $\overline{\mathcal{C}}_A\overline{\omega}_2{}^A{}_{A_1}$ on anti-ghosts $\overline{\mathcal{C}}_A$.

Given the specifics of gauge fixing procedure for the reducible
gauge symmetry with irreducible primary constraints and reducible
secondary ones, the complete BRST charge $Q$  reads
\begin{equation}\label{Qr}
Q=Q_{\text{min}}+\pi_\alpha P^\alpha + \pi_A\mathcal{P}^A +
\pi_{A_1}\mathcal{P}^{A_1} + \,\pi^{(1')}{\!}_{A_1}
\mathcal{P}^{(1')}{}^{A_1}\, .
\end{equation}
To accommodate all the gauge conditions, the gauge-fixing Fermion  $\Psi$ reads
\begin{equation}\label{Psired}
\begin{array}{c}
\displaystyle \Psi=\overline{C}_\alpha\, \chi^\alpha
+\overline{P}_\alpha\, \lambda^\alpha
+\overline{\mathcal{C}}_A\,  \omega_1{}^A{}_\alpha\, \mathcal{C}^\alpha
+\overline{\mathcal{P}}_A\, \lambda^A
+\overline{\mathcal{C}}_{A_1}\,  \omega_2{}^{A_1}{\!}_A\, \mathcal{C}^{A}
+\overline{\mathcal{P}}_{A_1}\, \lambda^{A_1}\\[2mm]
\displaystyle  +\,\overline{\mathcal{C}}{}^{(1')}{\!}_{A_1}\
\sigma_2{}^{A_1}{\!}_{A}\, \lambda^A +\overline{\mathcal{C}}{}_A\,
\overline{\omega}_2{}^{A}{}_{A_1}\ \lambda^{(1')}{}^{A_1}
+\frac{1}{2}\,\big(
\overline{\mathcal{C}}_{A_1}
\tau_2{}^{A_1}{\!}_{B_1}
\mathcal{P}^{(1')}{}^{B_1}
+
\pi{}_{A_1}
\varrho_2{}^{A_1}{\!}_{B_1}
\lambda^{(1')}{}^{B_1}
\big)\,.
\end{array}
\end{equation}
Given the complete BRST charge and gauge Fermion, the unitarizing
Hamiltonian is constructed by the usual rule (\ref{HPsiir}).

As we have seen in the previous section, the same Hamiltonian
action (\ref{SH}) with the involution relations
(\ref{irinv1})--(\ref{irinv5}) admits two different forms of gauge
symmetry: (i) the unfree gauge variations with first derivatives of
the gauge parameters (\ref{Hunfree1}), (\ref{Hunfree2}),
(\ref{HGPE}), and (ii) reducible gauge symmetry with the second
order derivatives of the unrestricted gauge parameters
(\ref{RedGS1}), (\ref{RedGS2}), (\ref{SofS1}), (\ref{SofS2}). These
two forms of the gauge symmetry are connected by the map of gauge
parameters (\ref{substtutionepsilon}). The map is not invertible if
the requirements of locality are respected. In this section, we have
presented two BRST complexes corresponding to these two forms of
gauge symmetry. Now, let us discuss the connection between these two
BRST complexes following the pattern (\ref{substtutionepsilon})
which links corresponding gauge symmetries. To avoid technical
complexities, we restrict consideration by simplified involution
relations
\begin{equation}\label{inv-simple}
\begin{array}{c}
\displaystyle \{T_\alpha,H\}=\tau_aV^a{}_\alpha, \quad \{\tau_a,H\}=T_\alpha V^\alpha{}_a\,,\\[2mm]
\displaystyle
\{T_\alpha,T_\beta\}=\{T_\alpha,\tau_a\}=\{\tau_a,\tau_b\}=0\,,
\end{array}
\end{equation}
with structure coefficients $V^a{}_\alpha$, $V^\alpha{}_a$ being
constants.

\noindent In this case,
$Q$ (\ref{Qir-nonmin}) reads
\begin{equation}\label{simpl-Qir-nonmin}
\displaystyle Q_{\texttt{unfree}}=T_\alpha C^\alpha+\tau_a
{C}^a+\pi_\alpha P^\alpha\,.
\end{equation}
This leads to the unitarizing Hamiltonian $H_\Psi$ (\ref{HPsiir}) of
the form
\begin{equation}\label{simpl-HPsiir}
\begin{array}{c}
\displaystyle H_\Psi^{\texttt{unfree}}=
H-\overline{P}_\alpha V^\alpha{}_a{C}^a-\overline{{P}}_aV^a{}_\alpha C^\alpha+T_\alpha\lambda^\alpha+\pi_\alpha\chi^\alpha\\[2mm]
\displaystyle+\,\overline{P}_\alpha P^\alpha+\overline{C}_\alpha\{\chi^\alpha,T_\beta\}C^\beta
+\overline{C}_\alpha\{\chi^\alpha,\tau_a\}{C}^a\,.
\end{array}
\end{equation}
Given the simplified algebra (\ref{inv-simple}), corresponding
involution relations for reducible constraints
$\widetilde{\tau}_\alpha$ (\ref{tildetau}) read
\begin{equation}\label{simplredinv}
\begin{array}{c}
\displaystyle \{T_\alpha, H\}=\widetilde{\tau}_\alpha, \quad \{\widetilde{\tau}_\alpha, H\}=T_\beta \widetilde{V}^\beta{}_\alpha\,,\\[2mm]
\displaystyle \{T_\alpha,T_\beta\}=\{T_\alpha,\widetilde{\tau}_\beta\}=
\{\widetilde{\tau}_\alpha,\widetilde{\tau}_\beta\}=0\,,
\end{array}
\end{equation}
\begin{equation}\label{simpl-Z-Z1}
\displaystyle \widetilde{\tau}_\alpha Z_{1}{}^{\alpha}{}_A=0\,, \quad Z_{1}{}^{\alpha}{}_A Z_{2}{}^{A}{}_{A_1}=0\,,
\end{equation}
with  $Z_1{}^{\alpha}{}_A$, $Z_2{}^{A}{}_{A_1}$ being
constants, and
${V}^{a}{}_{\alpha}Z_{1}{}^{\alpha}{}_{A}=0$,
$\widetilde{V}^{\alpha}{}_{\beta} = V^\alpha{}_a {V}^{a}{}_{\beta}$.

\noindent In this case, BRST charge (\ref{Qr}) reads
\begin{equation}\label{simpl-Q-red}
\begin{array}{c}
Q_{\texttt{reducible}}= T_\alpha\, C^\alpha
+\widetilde{\tau}_\alpha\, \mathcal{C}^\alpha +
\overline{\mathcal{P}}_\alpha\, Z_1{}^\alpha{}_A\, \mathcal{C}^A +
\overline{\mathcal{P}}_A \, Z_2{}^A{}_{A_1}\, \mathcal{C}^{A_1}\\[2mm]
\displaystyle +\,\pi_\alpha P^\alpha + \pi_A\mathcal{P}^A +
\pi_{A_1}\mathcal{P}^{A_1} + \,\pi^{(1')}{\!}_{A_1}
\mathcal{P}^{(1')}{}^{A_1}\,,
\end{array}
\end{equation}
and unitarizing Hamiltonian  (\ref{HPsiir}) has the form
\begin{equation}\label{simpl-HPsir}
\begin{array}{c}
\displaystyle H_{\Psi}^{\texttt{reducible}} =
H(q,p)
-\overline{P}_\beta\widetilde{V}^\beta{}_\alpha\mathcal{C}^\alpha
-\overline{\mathcal{P}}_\alpha\, C^\alpha\\[2mm]
\displaystyle +\,T_\alpha\lambda^\alpha
+\overline{\mathcal{P}}_\alpha Z_1{}^\alpha{}_A\lambda^A
+\overline{\mathcal{P}}_A Z_2{}^A{}_{A_1}\lambda^{A_1}
+\pi_A\,\overline{\omega}_2{}^{A}{}_{A_1}\lambda^{(1')}{}^{A_1}
+\overline{\mathcal{C}}{}_A\,\overline{\omega}_2{}^{A}{}_{A_1}\mathcal{P}^{(1')}{}^{A_1}
\\[2mm]
\displaystyle +\,\pi_\alpha\chi^\alpha
+\pi_A\,\omega_1{}^A{}_\alpha \mathcal{C^\alpha}
+\pi_{A_1}\,\omega_2{}^{A_1}{}_A \mathcal{C}^{A}
+\pi^{(1')}{\!}_{A_1}\,\sigma_2{}^{A_1}{\!}_{A} \lambda^A
+\overline{\mathcal{C}}{}^{(1')}{\!}_{A_1}\,
\sigma_2{}^{A_1}{\!}_{A} \mathcal{P}^A\\[2mm]
\displaystyle+\,\overline{C}_\alpha\{\chi^\alpha, T_\beta  \}\, C^\beta
+\overline{C}_\alpha\{\chi^\alpha , \widetilde{\tau}_\beta  \}\,\mathcal{C}^\beta
+\overline{\mathcal{C}}_A \,\omega_1{}^A{}_\alpha Z_1{}^\alpha{}_B\,
\mathcal{C}^B
+\overline{\mathcal{C}}_{A_1}\,
\omega_2{}^{A_1}{}_A  Z_2{}^A{}_{B_1}\mathcal{C}^{B_1}\\[2mm]
\displaystyle+\,\frac{1}{2}\,\pi_{A_1}\big( \tau_2{}^{A_1}{\!}_{B_1}
+\varrho_2{}^{A_1}{\!}_{B_1}\big)\mathcal{P}^{(1')}{}^{B_1}
+\overline{P}_\alpha P^\alpha
+\overline{\mathcal{P}}_A \mathcal{P}^A
+\overline{\mathcal{P}}_{A_1}\mathcal{P}^{A_1}\,.
\end{array}
\end{equation}

One can see, that even in the case of the simplest possible
involution relations (\ref{inv-simple}) leading to the secondary
constraints, two BRST complexes can be associated with the same
original classical action functional (\ref{SH}). One of them
corresponds to the unfree form of the gauge symmetry with the first
derivatives of gauge parameters, and another one is for the
reducible gauge symmetry with the second time derivatives of
unrestricted gauge parameters. The BRST charges
$Q_{\texttt{unfree}}$ and $Q_{\texttt{reducible}}$ begin with
different sets of secondary constraints -- $\tau_a$ and
$\widetilde{\tau}_\alpha$ -- and depend on the ghosts which differ
even by their numbers. The secondary constraints $\tau_a$ and
$\widetilde{\tau}_\alpha$ are linearly connected by relation
(\ref{tildetau}), where the linear differential operator
$V^a{}_\alpha$ has finite dimensional right kernel in the sense of
relation (\ref{KerG}). This results in the unfree gauge symmetry.
The left kernel is infinite  of $V^a{}_\alpha$, see (\ref{KERV}).
This fact results in the reducible gauge symmetry with the second
derivatives of the gauge parameters, and with the symmetry of
symmetry generated by the left kernel of $V^a{}_\alpha$. The connection between
these two symmetries is seen from the relation
(\ref{substtutionepsilon}). This relation shows that the structure
coefficient $V^a{}_\alpha$ can be absorbed by the gauge parameters
of the reducible transformations resolving, in a sense, the
equations for the gauge parameters of unfree transformations in
terms of the reducible parameters and their first derivatives. It is
a local map of gauge symmetries which does not have the local
inverse, in general. One can expect to  see the ghosts corresponding
to $\tau_a$ and $\widetilde{\tau}_\alpha$ to be connected by a
similar map. With this idea in mind, we are seeking for the
canonical transformation of the $Q_{\texttt{unfree}}$ such that
maps the ghosts $C^a$ assigned to the irreducible
constraints $\tau_a$ to the ghosts $\mathcal{C}^\alpha$
corresponding to the reducible constraints $\widetilde{\tau}_\alpha$. Clearly, no
canonical transformation can literally map the BRST charge
$Q_{\texttt{unfree}}$ to $Q_{\texttt{reducible}}$, or vice versa, as
they involve different number of ghosts, with different gradings.
The best possible way to see them connected by canonical
transformation is to identify the charges modulo trivial canonical
ghost pairs such that do not contribute to the BRST-cohomology
classes. For more details on the trivial pairs, one can consult the
textbook \cite{teitelboim1992quantization}.

Proceeding from the BRST complex connected with the unfree gauge
symmetry (\ref{simpl-Qir-nonmin}), we introduce the trivial pairs of
ghost variables to make the set of variables identical with that of
the complex for the reducible gauge symmetry (\ref{simpl-Q-red}),
\begin{equation}\label{cApB}
\displaystyle \{c^A,\overline{p}_B\}=\delta^A_B\,, \quad \text{gh}\,c^A=-\,\text{gh}\,\overline{p}_A=1\,, \quad \varepsilon(c^A)=\varepsilon(\overline{p}_A)=1\,;
\end{equation}
\begin{equation}\label{trivpairs2}
\displaystyle \{c^{A_1},\overline{p}_{B_1}\}=\delta^{A_1}_{B_1}\,,
\quad \text{gh}\,c^{A_1}=-\,\text{gh}\,\overline{p}_{A_1}=2\,, \quad
\varepsilon(c^{A_1})=\varepsilon(\overline{p}_{A_1})=0\,;
\end{equation}
\begin{equation}\label{trivpairs3}
\displaystyle
\{\mathfrak{c}^A,\overline{\mathfrak{p}}_B\}=\delta^A_B\,, \quad
\text{gh}\,\mathfrak{c}^A=-\,\text{gh}\,\overline{\mathfrak{p}}_A=2\,,
\quad
\varepsilon(\mathfrak{c}^A)=\varepsilon(\overline{\mathfrak{p}}_A)=0\,;
\end{equation}
\begin{equation}
\displaystyle \{\mathfrak{c}^{A_1},\overline{\mathfrak{p}}_{B_1}\}=\delta^{A_1}_{B_1}\,, \quad \text{gh}\,\mathfrak{c}^{A_1}=-\,\text{gh}\,\overline{\mathfrak{p}}_{A_1}=3\,, \quad \varepsilon(\mathfrak{c}^{A_1})=\varepsilon(\overline{\mathfrak{p}}_{A_1})=1\,,
\end{equation}
and corresponding non-minimal sector ghosts and Lagrange multipliers
\begin{equation}
\displaystyle \{\mathfrak{p}^A,\overline{\mathfrak{c}}{}_{B}\}=\delta^A_B\,, \quad \text{gh}\,\mathfrak{p}^A=-\,\text{gh}\,\overline{\mathfrak{c}}{}_{A}=2\,, \quad \varepsilon(\mathfrak{p}^A)=\varepsilon(\overline{\mathfrak{c}}{}_{A})=0\,;
\end{equation}
\begin{equation}
\displaystyle \{\lambda'^A,\pi'_B\}=\delta^A_B\,, \quad
\text{gh}\,\lambda'^A=-\,\text{gh}\,\pi'_A=1\,, \quad
\varepsilon(\lambda'^A)=\varepsilon(\pi'_A)=1\,;
\end{equation}
\begin{equation}
\displaystyle
\{\mathfrak{p}^{A_1},\overline{\mathfrak{c}}{}_{B_1}\}=\delta^{A_1}_{B_1}\,,
\quad
\text{gh}\,\mathfrak{p}^{A_1}=-\,\text{gh}\,\overline{\mathfrak{c}}{}_{A_1}=3\,,
\quad
\varepsilon(\mathfrak{p}^{A_1})=\varepsilon(\overline{\mathfrak{c}}{}_{A_1})=1\,;
\end{equation}
\begin{equation}\label{lambda'A1pi'B1}
\displaystyle \{\lambda'^{A_1},\pi'_{B_1}\}=\delta^{A_1}_{B_1}\,, \quad
\text{gh}\,\lambda'^{A_1}=-\,\text{gh}\,\pi'_{A_1}=2\,, \quad
\varepsilon(\lambda'^{A_1})=\varepsilon(\pi'_{A_1})=0\,.
\end{equation}
Now, let us extend the BRST charge  $Q_{\texttt{unfree}}$
(\ref{simpl-Qir-nonmin}) by the trivial pairs:
\begin{equation}\label{Q-unfree-ext}
Q_{\texttt{unfree}}^{\texttt{ext}}= T_\alpha C^\alpha +\tau_a C^a
+\pi_\alpha P^\alpha +\overline{p}_A\mathfrak{c}^A
+\overline{p}_{A_1} \mathfrak{c}^{A_1} +\pi'_A\mathfrak{p}^A
+\pi'_{A_1}\mathfrak{p}^{A_1}\,.
\end{equation}
Once the added trivial pairs are decoupled from the original
variables in the charge, they cannot contribute to the
BRST-cohomology classes. This means, the complex defined by the
original charge $Q_{\texttt{unfree}}$ (\ref{simpl-Qir-nonmin}) is
quasi-isomorphic to the above extension. The extended charge
(\ref{Q-unfree-ext}) is defined on the space of the same dimension
and with the same  distribution of the ghost number and
Grassmann-parity grading as the charge $Q_{\texttt{reducible}}$
(\ref{simpl-Q-red}) which corresponds to the reducible gauge
symmetry. Both charges are linear in the ghosts, so they can be
connected by  the linear canonical transformation of the ghosts and
Lagrange multipliers.

The distinctions between $Q_{\texttt{unfree}}^{\texttt{ext}}$ and
$Q_{\texttt{reducible}}$ concern the coefficients at the ghosts. The
charge $Q_{\texttt{unfree}}^{\texttt{ext}}$ includes the irreducible
secondary constraints $\tau_a$, while corresponding ghosts and trivial pairs decouple. The charge $Q_{\texttt{reducible}}$
(\ref{simpl-Q-red}) includes the reducible secondary constraints
$\widetilde{\tau}_\alpha$ connected with $\tau_a$ by the relation
(\ref{tildetau}). The momenta of corresponding ghosts
$\mathcal{C}^\alpha$ do not decouple from  the ghosts of higher
ghost numbers. So, the charge $Q_{\texttt{unfree}}^{\texttt{ext}}$
(\ref{Q-unfree-ext}) looks as if the ghosts in the charge
$Q_{\texttt{reducible}}$ (\ref{simpl-Q-red}) were reorganized  in
such a way that the corresponding secondary constraints were
explicitly split into irreducible ones and identical zeros, while
the null-vectors $Z_1{}^\alpha{}_A, Z_2{}^A{}_{A_1}$, being rectangular matrices, were split
into unit blocks and zero ones. Hence, to connect these two charges
we are seeking for the linear canonical transformation of the ghosts
and Lagrange multipliers such that corresponds to splitting of the
reducible constraints $\widetilde{\tau}_\alpha$ into irreducible
ones, i.e. $\tau_a$, and identical zeros. Below, we find the local
generating function\footnote{Locality of the generating function
does not necessarily mean that the canonical transformation is
local. For example, if the coordinates are connected by the
differential operators involved in the generating function, the
conjugate momenta are connected by the inverse operators. The latter
ones are non-local, in general.} of the canonical transformation
which connects the local charges
$Q_{\texttt{unfree}}^{\texttt{ext}}$ and $Q_{\texttt{reducible}}$ to
each other.

To find the linear canonical transformation connecting these two
alternative BRST charges, we introduce the local operators,
\begin{equation}\label{hatoperators}
\widehat{V}^\alpha{}_a\,,
\qquad
\widehat{Z}_1{}^A{}_\alpha\,,
\qquad
\widehat{Z}_2{}^{A_1}{\!}_A\,,
\end{equation}
subject to the conditions
\begin{equation}
\widehat{Z}_1{}^A{}_\alpha \widehat{V}^\alpha{}_a = 0\,, \qquad
\widehat{Z}_2{}^{A_1}{\!}_A\, \widehat{Z}_1{}^A{}_\alpha = 0\,.
\end{equation}
In this context, these operators are supposed field-independent much like $V^a{}_\alpha, Z_1{}^\alpha{}_A, Z_2{}^A{}_{A_1}$.
Introduce the operators
\begin{eqnarray} \label{Dab}
D^a{}_b &=& V^a{}_\alpha \widehat{V}^\alpha{}_b\,,\\\label{Dalphabeta}
\mathcal{D}^\alpha{}_\beta&=& \widehat{V}^\alpha{}_a\, V^a{}_\beta +
Z_1{}^\alpha{}_A\, \widehat{Z}_1{}^A{}_\beta\,,\\ \label{DAB}
\mathcal{D}^A{}{}_B&=& \widehat{Z}_1{}^A{}_\alpha\, Z_1{}^\alpha{}_B
+  Z_2{}^A{}_{A_1}\, \widehat{Z}_2{}^{A_1}{\!}_B\,,\\
\label{DA1B1} \mathcal{D}^{A_1}{}_{B_1} &=&
\widehat{Z}_2{}^{A_1}{\!}_A\, Z_2{}^A{}_{B_1}\,.
\end{eqnarray}
We assume, that all the operators above have at most finite
dimensional kernel. In this sense, the local operators
(\ref{hatoperators}) can be viewed as dual to $V^a{}_\alpha,
\,Z_1{}^\alpha{}_A,\, Z_2{}^A{}_{A_1}$, respectively. Given the
definitions, the operators (\ref{Dab}), (\ref{Dalphabeta}),
(\ref{DAB}), (\ref{DA1B1}) have the following inter-twinning
properties w.r.t. $V^a{}_\alpha,Z_{1}{}^\alpha{}_A, Z_2{}^A{}_{A_1}$ and corresponding dual operators:
\begin{equation}
\begin{array}{c}
V^a{}_\beta\mathcal{D}^\beta{}_\alpha=
D^a{}_bV^b{}_\alpha\,,
\qquad
\mathcal{D}^\alpha{}_\beta\widehat{V}^\beta{}_a=
\widehat{V}^\alpha{}_b D^b{}_a\,,
\\[2mm]
Z_1{}^\alpha{}_B\mathcal{D}^B{}{}_A=
\mathcal{D}^\alpha{}_\beta Z_1{}^\beta{}_A{}\,,
\qquad
 \widehat{Z}_1{}^A{}_\beta\mathcal{D}^\beta{}_\alpha=
\mathcal{D}^A{}_B\widehat{Z}_1{}^B{}_\alpha\,,
\\[2mm]
Z_2{}^A{}_{B_1}\mathcal{D}^{B_1}{}_{A_1} =
\mathcal{D}^A{}_B Z_2{}^B{}_{A_1}\,,
\qquad
\mathcal{D}^{A_1}{\!}_{B_1}\widehat{Z}_2{}^{B_1}{\!}_A =
\widehat{Z}_2{}^{A_1}{\!}_B\mathcal{D}^B{}{}_A\,.
\end{array}
\end{equation}
Hereinafter, let us denote as $(q,p)$ the canonical
pairs of ghosts and Lagrange multipliers  of
$Q_{\texttt{unfree}}^{\texttt{ext}}$  which are involved in the
change of variables. Corresponding canonical pairs of
$Q_{\texttt{reducible}}$  are denoted $(Q,P)$,
\begin{equation}\label{collect-not}
\begin{array}{l}
\displaystyle q=\{C^a,c^A,\mathfrak{c}^A,c^{A_1},\mathfrak{c}^{A_1},
\mathfrak{p}{}^A, \lambda'^A, \mathfrak{p}^{A_1},\lambda'^{A_1}\}\,,\\[2mm]
\displaystyle p=\{\overline{P}_a,\,\overline{p}_A,\overline{\mathfrak{p}}_A,
\overline{p}_{A_1},\overline{\mathfrak{p}}_{A_1}\,,
\overline{\mathfrak{c}}{}_{A}, \pi'_A, \overline{\mathfrak{c}}{}_{A_1},\pi'_{A_1}\},\\[2mm]
\displaystyle Q=\{\mathcal{C}^\alpha, \lambda^{(1')A_1},\mathcal{C}^A,\mathcal{P}^{(1')A_1},\mathcal{C}^{A_1},
\mathcal{P}^A, \lambda^A, \mathcal{P}^{A_1},\lambda^{A_1}\}\,,\\[2mm]
\displaystyle P=\{\overline{\mathcal{P}}_\alpha,\pi^{(1')}{}_{A_1}, \overline{\mathcal{P}}_A, \overline{\mathcal{C}}{}^{(1')}{}_{A_1},\overline{\mathcal{P}}_{A_1},
\overline{\mathcal{C}}{}_{A}, \pi_A,\overline{\mathcal{C}}_{A_1},\pi_{A_1}\}\,.
\end{array}
\end{equation}
Consider the point canonical transformation $(q,p)\,\leftrightarrow
\,(Q,P)$, with the generating function
\begin{equation}\label{F}
\begin{array}{c}
\mathcal{F}(Q,p)= \big(\overline{P}_a V^a{}_\alpha+\overline{p}_A\widehat{Z}_1{}^A {}_\alpha{}\big)\mathcal{C}^\alpha+\overline{p}_A Z_2{}^A{}_{A_1}\lambda^{(1')}{}^{A_1}+\big(\overline{\mathfrak{p}}_B \widehat{Z}_1{}^B {}_\alpha Z_1{}^\alpha{}_A
+\overline{p}_{A_1}\widehat{Z}_2{}^{A_1}{}_A\big)\mathcal{C}^A+\\[2mm]
\displaystyle +\,\big(\overline{\mathfrak{p}}_A Z_2{}^A{}_{A_1}-\overline{p}_{A_1}\big)\mathcal{P}^{(1')}{}^{A_1}
+\overline{\mathfrak{p}}_{A_1}\mathcal{D}^{A_1}{\!}_{B_1}\mathcal{C}^{B_1}+\\[2mm]
\displaystyle+\,\overline{\mathfrak{c}}{}_A\mathcal{D}^{A}{}_{B}\mathcal{P}^B
+\pi'_{A}\mathcal{D}^{A}{\!}_{B}\lambda^{B}
+\overline{\mathfrak{c}}{}_{A_1}\mathcal{D}^{A_1}{\!}_{B_1}\mathcal{P}^{B_1}
+\pi'_{A_1}\mathcal{D}^{A_1}{\!}_{B_1}\lambda^{B_1}\,.
\end{array}
\end{equation}
The  generating function defines the linear transformation of
coordinates and momenta by the rule
\begin{equation}\label{CanTrans}
\displaystyle P=\frac{\partial{}^r{\!}\mathcal{F}}{\partial Q}\,,
\qquad \displaystyle q=\frac{\partial{}^l{\!}\mathcal{F}}{\partial
p}\,.
\end{equation}
The relations above are local. This does not mean, however, the locality
 of the transformation as such. The matter is that expression of the coordinates $q$  is local in terms of $Q$, not vice versa.
 The momenta $P$ are the local linear combinations of $p$, while the local inverse transformation does not exist, in general. Therefore, neither map $(q,p)\mapsto (Q,P)$ is local, nor is the inverse one\footnote{Once the matrix $\frac{\partial^2F}{\partial Q\partial p}$  has a finite kernel, the transformation (\ref{CanTrans}) is not a canonical one, strictly speaking. More accurately, this transformation belongs to the class of the thick morphisms of $Q$-manifolds \cite{VoronovTh2017}.}.
 Besides the issue of locality, the problem has to be addressed of the possible finite dimensional kernel of the operators (\ref{Dab})--(\ref{DA1B1}) involved in the transformation (\ref{F}), (\ref{CanTrans}). The kernel of these operators primarily originates from the kernel of the structure coefficient at the completion function (\ref{KerG}). If one fixes certain element of the kernel, the operators  (\ref{Dab})--(\ref{DA1B1}) will become invertible, though the inverse ones are non-local. The example might be $D=\Delta$. The Laplace operator has kernel unless the boundary conditions are specified for the functions. Fixing boundary conditions, one can restrict the kernel, or make it trivial. For the latter class of functions, the Laplacian is invertible, though the inverse is non-local. The charge $Q_{\texttt{unfree}}^{\texttt{ext}}$
(\ref{Q-unfree-ext}) involves the secondary constraints $\tau_a$ that include fixed modular parameters $\Lambda$. Hence, the charge corresponds to the unique element of the kernel (\ref{KerG}).
With this regard, we assume that non-local inverse operators exist for all the operators (\ref{Dab})--(\ref{DA1B1}).
Making use of these inverse operators, one can explicitly find the canonical transformation from $(q,p)$ to $(Q,P)$, and back.
This transformation turns $Q_{\texttt{unfree}}^{\texttt{ext}}$ (\ref{Q-unfree-ext}) into $Q_{\texttt{reducible}}$ (\ref{simpl-Q-red}):
\begin{equation}\label{QQ}
Q_{\texttt{unfree}}^{\texttt{ext}}(q(Q,P), p(Q,P)) = Q_{\texttt{reducible}}(Q,P)\, .
\end{equation}
As one can see, the two local BRST charges constructed for the same Hamiltonian action (\ref{SH}) are connected by the non-local canonical transformation  (\ref{CanTrans}) with the local generating function (\ref{F}). Given the non-local nature of the transformation connecting the local charges, corresponding local BRST complexes are not quasi-isomorphic, in general.

 The most obvious evidence of inequivalence of these two BRST complexes concerns the irreducible secondary constraints $\tau_a$, being the Hamiltonian counterparts of completion functions (\ref{tau}). These quantities are BRST exact w.r.t. the charge $Q_{\texttt{unfree}}$ (\ref{simpl-Qir-nonmin}),
 \begin{equation}\label{exact-tau}
   \tau_a=\{ \overline{P}_a, Q_{\texttt{unfree}}\} \, .
 \end{equation}
The same quantities are BRST-closed w.r.t. the
charge $Q_{\texttt{reducible}}$ (\ref{simpl-Q-red}),
\begin{equation}\label{closed-tau}
  \{\tau_a, Q_{\texttt{reducible}}\}=0\, ,
\end{equation}
though they are not BRST-exact. Let us demonstrate the latter fact. Assume the local quantities $\psi_a$ exist, such that
 \begin{equation}\label{psi-a}
    \tau_a=\{\psi_a, Q_{\texttt{reducible}}\} \, .
 \end{equation}
Given the ghost number grading, and linearity of $Q_{\texttt{reducible}}$ (\ref{simpl-Q-red}), this would mean $\psi_a=\overline{\mathcal{P}}_\alpha\psi^\alpha{}_a$, with $\psi^\alpha{}_a$ being local ghost-independent differential operator. Substituting $\psi_a$ of this form into (\ref{psi-a}) and accounting for (\ref{tildetau}) we arrive at the relation
\begin{equation}\label{tau-T}
 \tau_a= \{\psi_a, Q_{\texttt{reducible}}\}=\widetilde{\tau}_\alpha\psi^\alpha{}_a= \tau_bV^b{}_\alpha\psi^\alpha{}_a\, .
\end{equation}
This means $V^b{}_\alpha\psi^\alpha{}_a=\delta_a^b$, while no local $\psi^\alpha{}_a$ can exist with this property, see (\ref{inverseV}).

Let us summarise the distinctions between the two BRST complexes. The complex related to the unfree form of the gauge symmetry treats the independent secondary constraints $\tau_a$ (\ref{Htau}), being the Hamiltonian counterparts of completion functions (\ref{tau}), as trivial quantities. Once they correspond to BRST co-boundary, they are not observables from the standpoint of this complex. The complex based on the reducible form of gauge symmetry treats the reducible secondary constraints $\widetilde{\tau}_\alpha$ (\ref{tildetau}) as trivial quantities. Even though  $\widetilde{\tau}_\alpha$ is proportional to $\tau_a$, the finite kernel of the coefficient $V^a{}_\alpha$ makes the difference. As the result, $\tau_a$ is not BRST-exact, due to the modes, which contribute to the kernel. These global modes are not fixed by $\widetilde{\tau}_\alpha$. They just conserve (as $\tau_a$ commutes on shell with the Hamiltonian), with the unfixed initial data.  Given these distinctions between the two complexes, different options are possible for physical interpretation of the dynamics. The complex based on the unfree form of the gauge symmetry better suits for studying dynamics with fixed modular parameters.
The global conserved quantities (\ref{J}), being defined on shell by the modular parameters, are BRST-trivial from the standpoint of this complex. No quantum fluctuations, or quantum transitions can be assigned to these quantities. The complex based on the reducible form of the gauge symmetry better suits for consideration with unfixed modular parameters. It does not reduce the classical dynamics to the level surface of the global conserved quantities (\ref{J}). These quantities are BRST closed, not BRST-exact w.r.t. the BRST charge (\ref{Qminr}), (\ref{Qr}). With this form of gauge symmetry, the BRST complex allows one to ask questions about spectrum of these global conserved quantities, their quantum fluctuations and transitions, at least in principle. The choice between these two inequivalent complexes depends on the reasons of physical interpretation, it cannot be motivated by formal (dis)advantages of any of these two formalisms --- both of them are self-consistent.

\section{Example: unfree and reducible gauge symmetry in the Hamiltonian formalism of linearized unimodular gravity}\label{Section5}
In this section, we exemplify all the general structures discussed in
the previous two sections by the model of linearized unimodular
gravity. Hamiltonian action (\ref{SH}) for this model reads:
\begin{equation}\label{SH-LUG}
\displaystyle S=\int d^4x\,\big(\Pi^{\alpha\beta}\dot{h}_{\alpha\beta}-H_T\big)\,, \quad H_T=H+\lambda^\alpha T_\alpha\,,
\end{equation}
\begin{equation}\label{HLUG}
\displaystyle H=\Pi_{\alpha\beta}\Pi^{\alpha\beta}-\frac{1}{2}\Pi^2-\frac{1}{4}\big(\partial_\alpha h_{\beta\gamma}\partial^\alpha h^{\beta\gamma}+\partial_\alpha h\partial^\alpha h-2\partial_\alpha h_{\beta\gamma}\partial^\beta h^{\alpha\gamma}\big)\,,
\end{equation}
\begin{equation}\label{TalphaLUG}
\displaystyle T_\alpha=-\,2\partial^\beta \Pi_{\beta\alpha}\,,
\end{equation}
where $\alpha,\beta,\gamma=1,2,3$, $\eta_{\alpha\beta}=-\,\delta_{\alpha\beta}$, $h=\eta^{\alpha\beta}h_{\alpha\beta}$, $\Pi=\eta_{\alpha\beta}\Pi^{\alpha\beta}$.

The requirement of stability \noindent (\ref{Inv1}) of the primary constraints (\ref{TalphaLUG}),
\begin{equation}
\displaystyle \{T_\alpha,H_T\}=-\,\partial_\alpha\tau\,,
\end{equation}
leads to a single secondary constraint
\begin{equation}\label{LUGtau}
\displaystyle \tau=\partial^\alpha\partial^\beta h_{\alpha\beta}-\partial_\alpha\partial^\alpha h-\Lambda\,, \quad \Lambda=const\,,
\end{equation}
and no tertiary constraints appear (cf. (\ref{Inv2})),
\begin{equation}\label{tauLUG}
\displaystyle \{\tau,H_T\}=-\,\partial^\alpha T_\alpha\,.
\end{equation}
The role of the differential operator with finite kernel $\Gamma_1{}^a{}_\alpha$
(see (\ref{Inv1})) is plaid by the partial derivative. Hence, the
kernel is one dimensional, being constituted by constants. The expression $\partial^\alpha\partial^\beta h_{\alpha\beta}-\partial_\alpha\partial^\alpha h$ in (\ref{LUGtau}) reduces on shell to half scalar curvature.  So, the relation (\ref{LUGtau}) is the phase space counterpart  of (\ref{TgLambda}) for linearized theory.
Involution relations (\ref{irinv1})--(\ref{irinv5}) for  the linearized unimodular gravity are of the form (\ref{inv-simple}),
\begin{equation}\label{inv-simple-LUG}
\begin{array}{c}
\displaystyle \{T_\alpha,H\}=-\,\partial_\alpha\tau\,, \quad \{\tau,H\}=-\,\partial^\alpha T_\alpha\,,\\[2mm]
\displaystyle \{T_\alpha,T_\beta\}=\{T_\alpha,\tau\}=\{\tau,\tau\}=0\,.
\end{array}
\end{equation}
Given the involution relations, the
unfree gauge symmetry transformations are constructed by the general
recipe (\ref{Hunfree1})--(\ref{HGPE}):
\begin{equation}\label{Hunfree1-LUG}
\displaystyle \delta_\epsilon h_{\alpha\beta}=\partial_\alpha\epsilon_\beta+\partial_\beta\epsilon_\alpha\,, \quad
\delta_\epsilon \Pi^{\alpha\beta}=-\,\partial^\alpha\partial^\beta\epsilon+\eta^{\alpha\beta}\partial_\gamma\partial^\gamma\epsilon\,,
\end{equation}
\begin{equation}\label{Hunfree2-LUG}
\delta_\epsilon\lambda^\alpha=\dot{\epsilon}^\alpha+\partial^\alpha\epsilon\,,
\end{equation}
\begin{equation}\label{HGPE-LUG}
\displaystyle \dot{\epsilon}+\partial_\alpha\epsilon^\alpha=0\,.
\end{equation}
Corresponding BRST charge (\ref{simpl-Qir-nonmin}) reads
\begin{equation}\label{simpl-Qir-nonmin-LUG}
\displaystyle Q_{\texttt{unfree}}=-\,2\partial^\beta\Pi_{\beta\alpha}C^\alpha+(\partial_\alpha\partial_\beta h^{\alpha\beta}-\partial_\alpha\partial^\alpha h-\Lambda)C+\pi_\alpha P^\alpha\,.
\end{equation}
This leads to the unitarizing Hamitonian (\ref{simpl-HPsiir}) of the form
\begin{equation}\label{simpl-HPsiir-LUG}
\begin{array}{c}
\displaystyle H_\Psi^{\texttt{unfree}}=
H-\overline{P}_\alpha \partial^\alpha{C}-\overline{{P}}\partial_\alpha C^\alpha+T_\alpha\lambda^\alpha+\pi_\alpha\chi^\alpha\\[2mm]
\displaystyle+\,\overline{P}_\alpha P^\alpha+\overline{C}_\alpha\{\chi^\alpha,T_\beta\}C^\beta
+\overline{C}_\alpha\{\chi^\alpha,\tau\}{C}\,,
\end{array}
\end{equation}
where $T_\alpha$ and $\tau$ are primary constraints (\ref{TalphaLUG}) and secondary constraint (\ref{LUGtau}), respectively, $\chi^\alpha$ are the admissible gauge conditions (\ref{chi}), and $H$ is defined by (\ref{HLUG}).

Reducible constraints $\widetilde{\tau}_\alpha$ are connected with $\tau$ (\ref{LUGtau}) by the relation (\ref{tildetau}),
\begin{equation}
\displaystyle \widetilde{\tau}_\alpha=-\,\partial_\alpha \tau\,,
\end{equation}
i.e. the structure function $V^a{}_\alpha$ is defined by
\begin{equation}\label{V-LUG}
\displaystyle f_aV^a{}_\alpha=-\,\partial_\alpha f\,, \quad V^a{}_\alpha f^\alpha=\partial_\alpha f^\alpha\,,
\end{equation}
and index $a$ takes a single value.
Given the algebra (\ref{inv-simple-LUG}), for reducible constraints
\begin{equation}\label{tildetauLUG}
\displaystyle \widetilde{\tau}_\alpha =
-\,\partial_\alpha\big(\partial^\beta\partial^\gamma
h_{\beta\gamma}-\partial_\beta\partial^\beta h\big)\,
\end{equation}
involution relations are of the form (\ref{simplredinv}), (\ref{simpl-Z-Z1}),
\begin{equation}\label{simplredinv-LUG}
\begin{array}{c}
\displaystyle \{T_\alpha, H\}=\widetilde{\tau}_\alpha, \quad \{\widetilde{\tau}_\alpha, H\}=\partial_\alpha\partial^\beta T_\beta\,,\\[2mm]
\displaystyle \{T_\alpha,T_\beta\}=\{T_\alpha,\widetilde{\tau}_\beta\}=
\{\widetilde{\tau}_\alpha,\widetilde{\tau}_\beta\}=0\,,
\end{array}
\end{equation}
\begin{equation}\label{simpl-Z-Z1-LUG}
\displaystyle \widetilde{\tau}_\alpha Z_{1}{}^{\alpha A}=0\,, \quad Z_{1}{}^{\alpha A} Z_{2\,A\,A_1}=0\,,
\end{equation}
where $Z_1{}^{\alpha A}$, $Z_{2\,A\,A_1}$ are defined by
\begin{equation}\label{Z1-LUG}
\displaystyle f_\alpha Z_1{}^{\alpha A}=-\,\varepsilon^{A\beta\alpha}\partial_\beta f_\alpha\,, \quad Z_1{}^{\alpha A}f_A=-\,\varepsilon^{\alpha\beta A}\partial_\beta f_A\,,
\end{equation}
\begin{equation}\label{Z2-LUG}
\displaystyle f^A Z_{2\,A\,A_1}=-\partial_Af^A\,, \quad Z_{2\,A\,A_1}f^{A_1}=\partial_A f\,,
\end{equation}
$A=1,2,3$, index $A_1$ takes a single value, and $\varepsilon^{A\beta\alpha}$ is fully antisymmetric, $\varepsilon^{123}=-\,\varepsilon_{123}=1$.

Reducible gauge transformations of the phase-space variables (\ref{RedGS1}) and Lagrange multipliers (\ref{RedGS2}) with $Z_1{}^{\alpha A}$ (\ref{Z1-LUG}) and $Z_{2AA_1}$ (\ref{Z2-LUG}) read
\begin{equation}\label{RedGS1-LUG}
\displaystyle \delta_\varepsilon h_{\alpha\beta}=\partial_\alpha(\dot{\varepsilon}_\beta+\varepsilon_{\beta\gamma A}\,\partial^\gamma \varepsilon^A)+
\partial_\beta(\dot{\varepsilon}_\alpha+\varepsilon_{\alpha\gamma A}\,\partial^\gamma \varepsilon^A)\,, \quad
\delta_\varepsilon \Pi^{\alpha\beta}=\partial^\alpha\partial^\beta\partial_\gamma \varepsilon^\gamma-\eta^{\alpha\beta}\partial_\gamma\partial^\gamma\partial_\lambda\varepsilon^\lambda\,,
\end{equation}
\begin{equation}\label{RedGS2-LUG}
\displaystyle \delta_\varepsilon \lambda^\alpha=\ddot{\varepsilon}^\alpha+\varepsilon^{\alpha\beta A}\partial_\beta\dot{\varepsilon}_A-\partial^\alpha\partial_\beta\varepsilon^\beta\,.
\end{equation}
Gauge symmetry transformations of the original gauge parameters $\varepsilon^\alpha$ and $\varepsilon^A$
\begin{equation}
\displaystyle \delta_\omega\varepsilon^\alpha=-\,\varepsilon^{\alpha\beta A}\partial_\beta\omega_A\,, \quad
\delta_\omega\varepsilon^A=\dot{\omega}^A-\partial^A\omega\,,
\end{equation}
are further reducible,
\begin{equation}
\displaystyle \delta_\eta \omega^A=\partial^A \eta\,, \quad \delta_\eta\omega=\dot{\eta}\,,
\end{equation}
see (\ref{SofS1}), (\ref{SofS2}) with $Z_1{}^{\alpha A}$ (\ref{Z1-LUG}) and $Z_{2AA_1}$ (\ref{Z2-LUG}). Note, that the parameters $\epsilon^\alpha$, $\epsilon$ of unfree gauge transformations (\ref{Hunfree1-LUG})--(\ref{Hunfree2-LUG}), (\ref{HGPE-LUG}) can be expressed in terms of $\varepsilon^\alpha$, $\varepsilon\,^A$ from (\ref{RedGS1-LUG})--(\ref{RedGS2-LUG}) (cf. (\ref{substtutionepsilon})),
\begin{equation}
\displaystyle \epsilon^\alpha=\dot{\varepsilon}^\alpha+\varepsilon^{\alpha\beta A}\partial_\beta\varepsilon_A\,, \quad
\epsilon=-\,\partial_\alpha\varepsilon^\alpha\,.
\end{equation}

The BRST charge (\ref{simpl-Q-red}), with $Z_1{}^{\alpha A}$ and $Z_{2AA_1}$ defined by (\ref{Z1-LUG}) and (\ref{Z2-LUG}), reads
\begin{equation}\label{simpl-Q-red-LUG}
\begin{array}{c}
Q_{\texttt{reducible}}= T_\alpha\, C^\alpha
+\widetilde{\tau}_\alpha\, \mathcal{C}^\alpha -
\overline{\mathcal{P}}_\alpha\,\varepsilon^{\alpha\beta A}\partial_\beta\,\mathcal{C}_A +
\overline{\mathcal{P}}_A\partial^A\mathcal{C}\\[2mm]
\displaystyle +\,\pi_\alpha P^\alpha + \pi_A\mathcal{P}^A +
\pi\mathcal{P} + \,\pi^{(1')}\mathcal{P}^{(1')}\,,
\end{array}
\end{equation}
where $T_\alpha$ and $\widetilde{\tau}_\alpha$ are primary constraints (\ref{TalphaLUG}) and secondary constraint (\ref{tildetauLUG}), respectively.

Local operators $\widehat{V}^\alpha{}_b$, $\widehat{Z}_{1A\,\alpha}$ and $\widehat{Z}_2{}^{A_1}{}^A$ (\ref{hatoperators}) read
\begin{equation}
\displaystyle f_\alpha \widehat{V}^\alpha{}_b  =
\partial^\alpha f_\alpha\,, \quad
\widehat{V}^\alpha{}_b f^b =
-\,\partial^\alpha f\,,
\end{equation}
\begin{equation}
\displaystyle f^A\widehat{Z}_{1A\,\alpha}=\varepsilon_{\alpha\beta A}\partial^\beta f^A\,, \quad \widehat{Z}_{1A\,\alpha}f^\alpha=\varepsilon_{A\beta\alpha}\partial^\beta f^\alpha\,,
\end{equation}
\begin{equation}
f_{A_1}\,\widehat{Z}_2{}^{A_1}{}^A =\partial^A f\,, \quad
\widehat{Z}_2{}^{A_1}{}^A f_A =-\,\partial^A f_A\,.
\end{equation}
Given gauge coefficients
\begin{equation}\label{k}
\begin{array}{c}
\displaystyle \omega_{1A\,\alpha} = \widehat{Z}_{1A\,\alpha}\,, \quad
\omega_2{}^{A_1A}= \sigma_2{}^{A_1A}= \widehat{Z}_2{}^{A_1A}\,, \quad
\overline{\omega}_{2A\,A_1}=Z_{2A\,A_1}\,,\\[2mm]
\displaystyle \tau_2{}^{A_1}{\!}_{B_1}= -\,2\delta^{A_1}_{B_1}\,,\quad
\varrho_2{}^{A_1}{\!}_{B_1}=0\,,
\end{array}
\end{equation}
the unitarizing Hamiltonian  (\ref{simpl-HPsir}) has the form
\begin{equation}\label{simpl-HPsir-LUG}
\begin{array}{c}
\displaystyle H_{\Psi}^{\texttt{reducible}} =
H-\overline{P}_\beta\,\partial^\beta \partial_\alpha\mathcal{C}^\alpha
-\overline{\mathcal{P}}_\alpha C^\alpha\\[2mm]
\displaystyle +\,T_\alpha\lambda^\alpha
-\overline{\mathcal{P}}_\alpha\,
\varepsilon^{\alpha\beta}{}^A\, \partial_\beta \lambda_A
+\overline{\mathcal{P}}_A\, \partial^A\lambda
+\pi_A\,\partial^A\lambda^{(1')}
+\overline{\mathcal{C}}{}_A\,
\partial^A \mathcal{P}^{(1')}
\\[2mm]
\displaystyle +\,\pi_\alpha\chi^\alpha
+\pi^A\,\varepsilon_{A\gamma\alpha }   \partial^\gamma \mathcal{C}^\alpha
-\pi\,\partial_A \mathcal{C}^{A}
-\pi^{(1')}{\!}\,\partial_A  \lambda^A
-\overline{\mathcal{C}}{}^{(1')}{\!}\,
\partial_A \mathcal{P}^A\\[2mm]
\displaystyle+\,\overline{C}_\alpha\{\chi^\alpha, T_\beta  \}\, C^\beta
+\overline{C}_\alpha\{\chi^\alpha , \widetilde{\tau}_\beta  \}\,\mathcal{C}^\beta
+\overline{\mathcal{C}}_A\,
\partial^A\,\partial_B
\mathcal{C}^B
+\overline{\mathcal{C}}_A
\Delta
\mathcal{C}^A
+\overline{\mathcal{C}}\Delta
\mathcal{C}\\[2mm]
\displaystyle-\,\pi\,\mathcal{P}^{(1')}{}
+\overline{P}_\alpha P^\alpha
+\overline{\mathcal{P}}_A \mathcal{P}^A
+\overline{\mathcal{P}}\mathcal{P}\,,
\end{array}
\end{equation}
where $T_\alpha$ and $\widetilde{\tau}_\alpha$ are primary constraints (\ref{TalphaLUG}) and secondary constraint (\ref{tildetauLUG}), respectively, $\chi^\alpha$ are the admissible gauge conditions (\ref{chi}),  $H$ is defined by (\ref{HLUG}), $\Delta=-\,\partial_\alpha\partial^\alpha$.

The operators (\ref{Dab})--(\ref{DA1B1}) have the form
\begin{equation}
D^a{}_b=V^a{}_\alpha \,\widehat{V}^\alpha{}_b{} = \Delta\,,
\end{equation}
\begin{equation}
\mathcal{D}^\alpha{}_\beta=\widehat{V}^\alpha{}_a{}\,V^a{}_\beta
+Z_1{}^\alpha{}^A{}\,\widehat{Z}_1{}_A{}_\beta=
\delta_\beta^\alpha\Delta\,,
\end{equation}
\begin{equation}
\mathcal{D}_A{}^B=
\widehat{Z}_1{}_A{}_\alpha\,Z_1{}^\alpha{}^B
+Z_2{}_A{}_{A_1}\,\widehat{Z}_2{}^{A_1}{}^B =
\delta^B_A\, \Delta\,,
\end{equation}
\begin{equation}
\mathcal{D}^{A_1}{}_{B_1}=
\widehat{Z}_2{}^{A_1}{}^A\,Z_2{}_A{}_{B_1}=\Delta\,,
\end{equation}
and the generating function (\ref{F}) reads
\begin{equation}\label{F-LUG}
\begin{array}{c}
\mathcal{F}= \overline{P}\,\partial_\alpha\mathcal{C}^\alpha+\overline{p}^A\varepsilon_{A\beta\alpha}\partial^\beta\mathcal{C}^\alpha+\overline{p}_A \partial^A\lambda^{(1')}
+\overline{\mathfrak{p}}_A\partial^A\partial_B\mathcal{C}^B
+\overline{\mathfrak{p}}_A\Delta\mathcal{C}^A
-\overline{p}\,\partial_A\mathcal{C}^A\\[2mm]
\displaystyle +\,\overline{\mathfrak{p}}_A\partial^A\mathcal{P}^{(1')}
-\overline{p}\,\mathcal{P}^{(1')}
+\overline{\mathfrak{p}}\Delta\mathcal{C}
+\overline{\mathfrak{c}}{}_A\Delta\mathcal{P}^A
+\pi'_{A}\Delta\lambda^{A}
+\overline{\mathfrak{c}}\Delta\mathcal{P}
+\pi'\Delta\lambda\,.
\end{array}
\end{equation}
One can demonstrate, that the extended BRST charge (\ref{Q-unfree-ext}),
\begin{equation}\label{Q-unfree-ext-LUG}
Q_{\texttt{unfree}}^{\texttt{ext}}= -\,2\partial^\beta\Pi_{\beta\alpha}C^\alpha+(\partial_\alpha\partial_\beta h^{\beta\alpha}-\partial_\alpha\partial^\alpha h-\Lambda)C+\pi_\alpha P^\alpha +\overline{p}_A\mathfrak{c}^A
+\overline{p}\,\mathfrak{c} +\pi'_A\mathfrak{p}^A
+\pi'\mathfrak{p}\,,
\end{equation}
under the canonical transformation (\ref{CanTrans}) with the generating function $\mathcal{F}$ (\ref{F-LUG}) turns into $Q_\texttt{reducible}$ (\ref{simpl-Q-red-LUG}) (see (\ref{QQ})). Also, under such a canonical transformation, the extended $H_\Psi^{\texttt{unfree}}$ (\ref{simpl-HPsiir-LUG}),
\begin{equation}
\begin{array}{c}
\displaystyle H_{\Psi}^{\texttt{unfree\,ext}}=
H-\overline{P}_\alpha \partial^\alpha{C}-\overline{{P}}\partial_\alpha C^\alpha+T_\alpha\lambda^\alpha+\pi_\alpha\chi^\alpha\\[2mm]
\displaystyle+\,\overline{P}_\alpha P^\alpha+\overline{C}_\alpha\{\chi^\alpha,T_\beta\}C^\beta
+\overline{C}_\alpha\{\chi^\alpha,\tau\}{C}\\[2mm]
+\ \overline{p}_A\lambda'^A +\pi'_A\Delta c^A
+\overline{\mathfrak{p}}_A\mathfrak{p}^A
+\overline{\mathfrak{c}}{}_A \Delta\mathfrak{c}^A\\[2mm]
\ +\ \overline{p}\,\lambda'+\pi'\Delta c
+\overline{\mathfrak{p}}\mathfrak{p}
+\overline{\mathfrak{c}}\Delta\mathfrak{c}
-\overline{p}^A\varepsilon_{A\beta\alpha}\partial^\beta C^\alpha\,,
\end{array}
\end{equation}
being obtained by the appropriate choice of extended gauge Fermion (\ref{irredPsi})
\begin{equation}
\displaystyle \Psi_{\texttt{unfree}}^{\texttt{ext}}=
\overline{C}_\alpha\chi^\alpha
+\overline{P}_\alpha\lambda^\alpha
+\overline{\mathfrak{c}}_A\Delta c^A
+\overline{\mathfrak{p}}_A\lambda'^A
+ \overline{\mathfrak{c}}\Delta c
+\overline{\mathfrak{p}}\,\lambda'
-\overline{\mathfrak{p}}^A\varepsilon_{A\beta\alpha}\partial^\beta C^\alpha\,,
\end{equation}
turns into $H_{\Psi}^{\texttt{reducible}}$ (\ref{simpl-HPsir-LUG}), i.e.
\begin{equation}
H_{\Psi}^{\texttt{unfree\,ext}}=
H_{\Psi}^{\texttt{reducible}}\,.
\end{equation}

One can see, that
\begin{equation}
\displaystyle \tau=\{\overline{P}, Q_\texttt{unfree}\}\,,
\end{equation}
where $Q_\texttt{unfree}$ is defined by (\ref{simpl-Qir-nonmin-LUG}) (cf. (\ref{exact-tau})). In the same time, the irreducible secondary constraint can be represented as
\begin{equation}
\displaystyle \tau=\{\psi, Q_\texttt{reducible}\}\,,
\end{equation}
where $Q_\texttt{reducible}$ is defined by (\ref{simpl-Q-red-LUG}), and $\psi=\overline{\mathcal{P}}_\alpha\psi^\alpha$ (cf. (\ref{psi-a})). So, like in (\ref{tau-T}),
\begin{equation}
\displaystyle \tau=\widetilde{\tau}_\alpha\psi^\alpha=-\,\psi^\alpha\partial_\alpha\tau\,,
\end{equation}
i.e. $\psi^\alpha =\partial^\alpha\Delta^{-1}$ is non-local operator being inverse to partial derivative.
So, from the standpoint of the BRST charge (\ref{simpl-HPsir-LUG}), the scalar curvature, being represented by the independent secondary constraint $\tau$, is  BRST-exact.
Hence, the cosmological constant is a trivial quantity from the standpoint of this BRST-cohomology.
The BRST charge (\ref{simpl-Q-red-LUG}) defines another BRST complex, where $\tau$ is closed, while it is not BRST-exact.
Therefore, the cosmological constant is a physical quantity from the standpoint of this BRST complex.
The choice between these complexes depends on the problem setup.
If one sticks to the fixed asymptotic of the fields (i.e. the fields are supposed vanishing at infinity), the first complex is the option to choose.
If various asymptotics are considered admissible for the fields, one has to opt for the second complex, with unfixed $\Lambda$.

\section{Conclusion}

In this article, we consider the phenomenon of unfree gauge symmetry in constrained Hamiltonian formalism, and study its connection with the alternative description of the same system in terms of reducible gauge symmetry with unrestricted gauge parameters. In Section \ref{Section2} of this article, we remind the fact (which has been first noted in Ref. \cite{KAPARULIN2019114735}) that unfree gauge symmetry implies certain modification of the second Noether theorem (\ref{GI}), (\ref{KerG}) which accounts for the equations imposed onto the gauge parameters. This modification leads to existence of the global conserved quantities (\ref{tau}), (\ref{J}) whose on shell values are defined not by the initial data at the Cauchy hyper-surface but at the lower dimensional subset of the space-time, or even at one point. The best known example of the conserved quantity of this sort is the cosmological constant of unimodular gravity.  The cosmological constant enjoys the same status in various recent modifications of unimodular gravity, all related with restrictions on gauge parameters, see   \cite{BARVINSKY201759}, \cite{Barvinsky-Kolganov}. In the model with the volume preserving diffeomorphism recently proposed in the article \cite{Jirousek:2020vhy}, the role of the corresponding global conserved quantity  is plaid by the Newton constant. In the article \cite{Abakumova:2020ajc}, a series of the global conserved quantities is found for the higher spin analogue of the linearized unimodular gravity proposed in the article \cite{SKVORTSOV2008301}. In the present article, the new global conserved quantities of this type are found in the second example of Section \ref{Section2} for the Maxwell-like higher spin theories proposed in the article \cite{Campoleoni2013}.  As we demonstrate, these global conserved quantities are systematically deduced from the modification of the second Noether theorem (\ref{GI}), (\ref{KerG}) for the unfree gauge symmetry. In Section \ref{Section2}, we also demonstrate that any action, being invariant under the unfree gauge variation, always admits another form of gauge transformations with higher derivatives of unrestricted gauge parameters. This gauge symmetry can be reducible, even though the unfree symmetry of the same action is irreducible. This alternative form of the gauge symmetry does not lead to any modification of Noether identities. In Section \ref{Section3}, we work out  constrained Hamiltonian formalism describing both types of the gauge symmetry and clarify connection between them. The unfree gauge symmetry corresponds to the irreducible set of the secondary constraints, while the equations restricting the gauge parameters are defined by certain structure functions of the involution relations (\ref{Inv1}). The gauge symmetry  with unrestricted gauge parameters is generated by reducible set of secondary constraints. The reducible gauge transformations (\ref{RedGS1}), (\ref{RedGS2}), (\ref{HUnfreevariation}) involve the higher time derivatives of the gauge parameters.
The connection between these two forms of gauge symmetry is provided by the relation (\ref{substtutionepsilon}), which expresses the parameters of unfree gauge symmetry in terms of the parameters of reducible gauge transformations and their time derivatives.  This map of the gauge parameters is not invertible in local terms, in general. It demonstrates that the unfree gauge symmetry is stronger, in the sense that it can gauge out the quantities such that can be viewed non-trivial w.r.t. the reducible counterpart. In Section \ref{Section4}, we work out Hamiltonian BRST formalism both for the unfree and reducible form of gauge symmetry. Upon completion of the phase space of the first BRST theory by  certain set of BRST-trivial canonical pairs, it turns out connected with the formalism of the reducible gauge symmetry by the canonical transformation. The generating function (\ref{F}) is local of the transformation. The transformation (\ref{CanTrans}), as such, is not necessarily local, however. Hence, these two local BRST complexes are not necessarily quasi-isomorphic, in the sense that their cohomology groups can be different. In particular, the global conserved quantities (\ref{tau}), (\ref{J}) correspond to the BRST co-boundary for the first complex, while they are non-trivial co-cycles from the standpoint of the complex associated to the reducible gauge symmetry. Both forms of the gauge symmetry --- the unfree transformations and the reducible ones --- are self-consistent, though inequivalent at the level of corresponding BRST complexes. The choice between them depends on the physical problem setting. If the asymptotics is supposed fixed of the fields, and the initial data cannot vary for the global conserved quantities, the complex for unfree gauge symmetry is the option to choose. If the initial data are not fixed for the global conserved quantities, one will have to opt for the complex associated to the reducible gauge symmetry.
\subsection*{Note added.}When the work has been finalized, the article   \cite{Elfimov:2021aok} has appeared where the algebra is studied of the lower conserved quantities whose on-shell values are defined by the data on the lower dimension surfaces. These quantities seem similar to the global conserved quantities studied in this article, though no relation is immediately seen in \cite{Elfimov:2021aok} with the unfree gauge symmetry.
\subsection*{Acknowledgments.} The work of V. A. Abakumova and S. L. Lyakhovich is supported by the
Foundation for the Advancement of Theoretical Physics and
Mathematics ``BASIS", and by a government task of the Ministry of Science and Higher Education of the Russian Federation, Project No. 0721-2020-0033.

\end{document}